\documentclass[reprint,amsmath,amssymb,aps]{revtex4-2}

\usepackage{graphicx}
\usepackage{subcaption}
\usepackage{placeins}
\usepackage{float}
\usepackage{xcolor}
\usepackage{dcolumn}
\usepackage{bm}
\usepackage{xcolor}

\begin{document}

\preprint{APS/123-QED}

\title{Axial $w$-modes of anisotropic neutron stars}

\author{Sushovan Mondal}
\email{smondal@imsc.res.in}
\affiliation{Department of Physics, Indian Institute of Technology Bombay, Mumbai 400076, India.}
\affiliation{The Institute of Mathematical Sciences, HBNI, Taramani, Chennai 600113, India.}


\date{\today}

\begin{abstract}
	We investigate the axial $w$-mode oscillations of anisotropic neutron stars. Equilibrium stellar configurations are constructed using two realistic equations of state, BSk21 and SLy4, together with two prescriptions for pressure anisotropy, namely the Horvat ansatz and the Bowers--Liang ansatz. Only neutron-star models satisfying the stability and physical acceptability criteria are considered. On these backgrounds, the complex axial $w$-mode frequencies are computed by solving the linearized perturbation equations using a continued-fraction method. We find that, for each fixed anisotropy strength, the axial $w$-mode frequency decreases monotonically with increasing stellar mass along the stable branch, with its magnitude depending on both the equation of state and the nature of the anisotropy. At relatively low stellar masses, configurations with dominant radial pressure ($p_r>p_t$) exhibit higher frequencies than those with dominant tangential pressure, whereas toward the upper end of the stable branch this ordering is reversed, and configurations with $p_t>p_r$ attain higher frequencies at the same mass. When expressed as a function of compactness, the axial $w$-mode frequency displays an approximately linear dependence on $M/R$, with anisotropy modifying both the slope and the intercept of the relation. The Bowers--Liang prescription produces a wider spread in the frequency values compared to the Horvat ansatz. We also analyze the damping times associated with the axial $w$-modes and find that they increase with stellar mass, with a rapid rise toward the upper end of the stable branch. At a fixed mass, increasing the tangential pressure relative to the radial pressure leads to shorter damping times, while configurations with dominant radial pressure exhibit longer damping times. The sensitivity of the damping time to anisotropy is more pronounced for more compact stars, and the Bowers--Liang prescription yields systematically larger damping times than the Horvat ansatz. Finally, motivated by the numerical results, we provide empirical expressions for the axial $w$-mode frequency and damping time as functions of stellar compactness and anisotropy strength. These relations accurately capture the behavior of the axial $w$-mode spectrum within the physically allowed parameter space considered in this study.
\end{abstract}

\maketitle

\section{\label{intro} Introduction}

Neutron stars are among the most fascinating astrophysical objects, not only because they contain a mass greater than that of the Sun within a radius comparable to that of a city (about $10$ km), but also because they serve as natural laboratories for exploring physical phenomena under extreme conditions that are inaccessible in terrestrial settings. The theoretical conception of a star composed of extremely dense nuclear matter was first proposed by Lev Landau in 1932 \cite{Yakovlev:2012rd, Landau:1932uwv}. This was followed by a number of seminal theoretical studies that laid the foundations for our present understanding of neutron stars \cite{Baade:1934onh,Baade:1934zex,Tolman:1939jz,Oppenheimer:1939ne}. 

While the theoretical understanding of neutron stars was gradually developing, the discovery of pulsating radio sources in 1967 by Jocelyn Bell Burnell and Antony Hewish \cite{Hewish:1968bj}, later identified as rotating neutron stars and named pulsars, firmly established neutron stars as real astrophysical objects. Since then, more than $3000$ pulsars have been detected across various electromagnetic wavelengths, making them invaluable tools for probing fundamental physics under extreme conditions (for the most up-to-date count, see version~2.7.0 of \cite{Manchester:2005atnf}).
 
As neutron stars are extremely compact objects, general relativity plays a crucial role in describing their structure. Consequently, in addition to conventional observations across a wide range of electromagnetic wavelengths, gravitational waves (GWs) emitted by these sources carry a wealth of information that can be used to probe their internal properties. The direct detection of GWs about a decade ago by the LIGO--Virgo Collaboration \cite{LIGOScientific:2016aoc} opened a new observational window onto the Universe. Following this first detection, the LIGO--Virgo Collaboration observed gravitational waves from a binary neutron star merger in 2017 \cite{Abbott:2017mm}, marking the beginning of a new era in neutron star astrophysics.

Neutron stars can generate GWs through several mechanisms \cite{Lasky:2015uia}, among which nonradial oscillations are particularly promising sources. Phenomena such as pulsar glitches \cite{Haskell:2023exo,Haskell:2021ljd}, nonspherical supernova explosions \cite{Kumar:2024jky}, and the coalescence of neutron stars \cite{Benhar:1998au,Bauswein:2015vxa} are some of the possible triggers of such nonradial oscillations. The nonradial oscillation modes of neutron stars were first calculated by Thorne and Campolattaro \cite{thorne_campolattaro_1967}. Subsequently, a large number of studies have investigated these modes under various physical conditions, including superfluidity \cite{Gualtieri:2014lsa,Andersson:2002jd}, the presence of density discontinuities \cite{Sotani:2001bb,Miniutti:2002bh}, and the effects of inter-nucleon interactions \cite{Kunjipurayil:2022zah}. The impact of dark matter \cite{Dey:2024vsw,Sotani:2025lzy} as well as modified theories of gravity \cite{Blazquez-Salcedo:2021exm} on neutron star oscillation modes has also been explored. Most of these studies have primarily focused on the even-parity, or polar, fluid oscillation modes of neutron stars, such as the $f$-, $p$-, and $g$- modes.

In addition to the even-parity, or polar modes, neutron stars also admit odd-parity, or axial, oscillation modes. A distinguished subclass of axial oscillations, known as axial $w$-modes, consists of purely spacetime modes that
do not couple to fluid perturbations and have no Newtonian counterpart, in contrast to fluid axial modes such as $r$-modes, which are associated with horizontal fluid motions. Unlike fluid polar modes, axial $w$-modes do not couple to fluid perturbations and are purely spacetime oscillations, with no Newtonian counterpart. It is worth noting that $w$-modes also exist in the polar sector, although polar $w$-modes couple weakly to the fluid, in contrast to their axial counterparts. These modes are characterized by very high frequencies (of the order of $10$--$20$ kHz) and extremely short damping times (typically of the order of microseconds). Owing to these distinctive features, $w$-modes carry unique information about the stellar compactness and the equation of state (EOS), complementing the information obtained from fluid modes. Indeed, \citet{Benhar:1998au} demonstrated that axial $w$-mode frequencies are sensitive to the stiffness of the EOS, thereby encoding information about both the structure of neutron star matter and the underlying hadronic interactions.

$w$-modes of pulsating relativistic stars were first identified by \cite{Kokkotas:1986} and further studied in \cite{Kokkotas:1992}. The initial study of axial modes was carried out by \citet{chandra_1991_b}. Subsequently, \citet{Kokkotas:1994ax} computed the same modes using a different numerical approach and obtained results consistent with those of \citet{chandra_1991_b}. Later studies calculated axial $w$-modes for a wide range of EOSs (see \cite{Kokkotas:1999bd} for a comprehensive review). The imprint of different EOSs on the axial $w$-mode spectrum was quantitatively investigated by \citet{Benhar:1998au}, while \citet{Tsui:2004qd} showed that the axial $w$-mode frequencies obey nearly EOS-independent (universal) scaling relations when expressed in appropriate dimensionless form. Axial $w$-modes for neutron stars with realistic EOSs, including twin-star configurations, were studied by \citet{Chatterjee:2009w}. Beyond general relativity, axial $w$-modes of neutron stars have also been explored in alternative theories of gravity, such as scalar--tensor theories \cite{Sotani:2005,AltahaMotahar:2018}, Horndeski gravity \cite{BlazquezSalcedo:2018}, and $R^2$ gravity \cite{Blazquez-Salcedo:2018qyy}.
 
All of the studies discussed above have assumed that the pressure inside a neutron star is isotropic, i.e., the pressure is the same in all three spatial directions in hydrostatic equilibrium. However, the possibility of anisotropic pressure inside neutron stars cannot be ruled out. The concept of an anisotropic fluid, characterized by different pressures in the radial and tangential directions, was first introduced by \citet{Lemaitre:1933gd} in the context of cosmology. Subsequently, \citet{Bowers:1974tgi} investigated the equilibrium configurations of relativistic compact stars with pressure anisotropy and demonstrated that local anisotropy can significantly affect observable stellar properties such as the total mass and surface redshift.

Pressure anisotropy inside neutron stars may arise due to several physical mechanisms, such as superfluidity \cite{Ruderman:1972aj,Hoffberg1970}, pion condensation \cite{Sawyer:1972cq}, and skyrmionic interactions \cite{nelmes:2012}. The presence of viscosity can also induce local anisotropy in compact stars, including neutron stars \cite{barreto1992equation,barreto1993exploding}. In addition, \citet{Yazadjiev:2011ks} modeled magnetars in general relativity using a nonperturbative approach, where the stellar matter was treated as anisotropic due to the presence of strong magnetic fields. Elastic effects in compact stars, such as neutron stars, can likewise be described in terms of local anisotropy \cite{Karlovini:2002fc,alho2022compact}. Further discussions on anisotropic pressure in self-gravitating systems, including neutron stars, can be found in the review by Herrera and Santos \cite{Herrera:1997plx} and references therein.

The primary motivation of this work is to compute the axial quasinormal $w$-modes of anisotropic neutron stars. In recent studies, we have calculated the polar $f$-modes of anisotropic neutron stars \cite{Mondal:2023wwo} as well as anisotropic quark stars \cite{Mondal:2025ixk}. Polar modes of anisotropic neutron stars were also independently studied by \citet{Lau:2024oik}. Although these studies approached the problem using different formalisms, they were found to be in agreement for the $f$-mode oscillation properties. More recently, \citet{Yu:2026vbd} computed the polar $w$-modes of anisotropic neutron stars.

In the present work, we derive the governing equations for the axial $w$-modes of anisotropic neutron stars and compute their complex frequencies. To solve these equations and obtain the corresponding oscillation frequencies and damping times, we consider two realistic nuclear equations of state, namely BSk21 \cite{Potekhin:2013qqa} and SLy4 \cite{Douchin:2001sv,Chabanat:1998}, which relate the radial pressure to the energy density. To model pressure anisotropy, we adopt the ansatzes proposed by \citet{Horvat:2010xf} and \citet{Bowers:1974tgi}, both of which are widely used in the literature. The numerically computed results for the axial $w$-mode frequencies and the associated damping times are analyzed for different values of the anisotropy strength. In addition, we derive approximate analytical expressions for both the frequency and the damping time as functions of the stellar mass, radius, and anisotropic strength parameter. These relations may serve as useful tools for future studies of gravitational-wave signatures from compact objects.

This paper is organized as follows. In Sec.~\ref{sec:equilib}, we discuss the equilibrium structure of anisotropic neutron stars. In Sec.~\ref{sec:axial_eq_der}, we derive the governing equations for the axial $w$-modes of anisotropic neutron stars. In Sec.~\ref{sec:numerical_results}, we present the numerical results for the axial $w$-mode frequency and damping time, and derive approximate analytical expressions for these quantities as functions of the stellar mass, radius, and anisotropic strength. Finally, in Sec.~\ref{summ_con}, we summarize our findings and outline potential directions for future research. Throughout this work, we use geometrized units in which $G = c = 1$, where $G$ is the gravitational constant and $c$ is the speed of light in vacuum, unless otherwise stated.

\section{Equilibrium anisotropic configurations of neutron stars in general relativity}
\label{sec:equilib}

We consider a static, spherically symmetric compact star in general relativity, described by the line element
\begin{equation}
	ds^2 = -e^{\nu(r)} dt^2 + e^{\lambda(r)} dr^2 + r^2\left(d\theta^2 + \sin^2\theta\, d\phi^2\right),
\end{equation}
where $\nu(r)$ and $\lambda(r)$ depend only on the radial coordinate $r$.
For an anisotropic fluid, the energy-momentum tensor can be written as
\begin{equation}
	T^{\beta}{}_{\alpha}=(\rho+p_t)\,u^{\beta}u_{\alpha}+p_t\,\delta^{\beta}{}_{\alpha}+(p_r-p_t)\,s^{\beta}s_{\alpha},
\end{equation}
where $\rho$ is the energy density, $p_r$ and $p_t$ are the radial and tangential pressures, respectively, $u^\alpha$ is the fluid four-velocity, and $s^\alpha$ is the unit spacelike radial vector. For the metric given above,
\begin{equation}\label{uands}
	u^\alpha = \left(e^{-\nu/2},0,0,0\right),\qquad
	s^\alpha = \left(0,e^{-\lambda/2},0,0\right),
\end{equation}
which satisfy
\begin{subequations}\label{usnorm}
	\begin{align}
		u^\alpha u_\alpha &= -1,\\
		s^\alpha s_\alpha &= 1,\\
		u^\alpha s_\alpha &= 0.
	\end{align}
\end{subequations}
The nonvanishing components of $T^{\beta}{}_{\alpha}$ are therefore given by
$\mathrm{diag}(-\rho,\,p_r,\,p_t,\,p_t)$.

The spacetime geometry is determined by the Einstein equations,
\begin{equation}
	G^{\beta}{}_{\alpha}=8\pi\,T^{\beta}{}_{\alpha},
\end{equation}
which, for the above metric and stress-energy tensor, yield
\begin{align}
	e^{-\lambda}\left(\frac{\lambda'}{r}-\frac{1}{r^2}\right)+\frac{1}{r^2} &= 8\pi \rho, \\
	e^{-\lambda}\left(\frac{\nu'}{r}+\frac{1}{r^2}\right)-\frac{1}{r^2} &= 8\pi p_r, \\
	\frac{1}{2}e^{-\lambda}\left(\nu''-\frac{1}{2}\nu'\lambda'+\frac{1}{2}{\nu'}^2+\frac{\nu'-\lambda'}{r}\right) &= 8\pi p_t ,
\end{align}
where a prime denotes differentiation with respect to $r$.
Combining these equations, one obtains the hydrostatic equilibrium condition in the presence of pressure anisotropy (in units $G=c=1$),
\begin{equation}
	p_r' = -\frac{\nu'}{2}\,(\rho+p_r) + \frac{2\chi}{r},
\end{equation}
where $\chi \equiv p_t-p_r$ measures the local anisotropy.
It is convenient to introduce the mass function $m(r)$ through
\begin{equation}
	e^{-\lambda(r)} = 1-\frac{2m(r)}{r},\qquad
	m(r)=4\pi\int_{0}^{r}\rho(r')\,{r'}^{\,2}\,dr',
\end{equation}
which represents the gravitational mass enclosed within radius $r$.
Using this relation, the equilibrium equation can be recast into the modified Tolman--Oppenheimer--Volkoff (TOV) form,
\begin{equation}\label{eq:modTOV}
	p_r' = -\frac{(\rho+p_r)\left[m(r)+4\pi p_r r^3\right]}{r\left[r-2m(r)\right]}+\frac{2\chi}{r}.
\end{equation}
Thus, the stellar structure is completely specified once one provides (i) an equation of state $p_r(\rho)$ for neutron-star matter and (ii) a prescription for $\chi(r)$ (or, equivalently, for $p_t$).

For the boundary conditions, we impose regularity at the center, $m(0)=0$ and finite $p_r(0)\equiv p_c$, and define the stellar surface $r=R$ by the vanishing of the radial pressure, $p_r(R)=0$.
For standard neutron-star equations of state, this condition also implies $\rho(R)=0$.
Finally, the interior solution must match smoothly to the exterior Schwarzschild spacetime at $r=R$, which yields
\begin{equation}
	e^{\nu(R)}=e^{-\lambda(R)} = 1-\frac{2M}{R},\qquad M \equiv m(R).
\end{equation}

\subsection{Description of equations of state of neutron stars and anisotropy}

To construct equilibrium models of neutron stars, we solve the hydrostatic equilibrium equation, namely the modified Tolman--Oppenheimer--Volkoff (TOV) equation given in Eq.~(\ref{eq:modTOV}). This requires specifying an equation of state (EOS) relating the radial pressure $p_r$ to the energy density $\rho$, together with a prescription for the pressure anisotropy parameter $\chi$.

For the EOS, we consider two realistic nuclear models, BSk21 \cite{Potekhin:2013qqa} and SLy4 \cite{Douchin:2001sv,Chabanat:1998}, which are widely used in neutron-star structure calculations. To incorporate pressure anisotropy, we adopt two phenomenological ansatzes for $\chi$, both of which have been extensively employed in the literature.

The first ansatz, originally proposed by \citet{Horvat:2010xf}, is given by
\begin{equation}
	\chi = \tau_1\, p_r\, \mu ,
	\label{quasiparam}
\end{equation}
where $\mu = 2m/r$ denotes the local compactness and $\tau_1$ controls the strength of the anisotropy. This form has several appealing features. First, the anisotropy vanishes at the stellar center, since $\mu \sim r^2$ as $r \to 0$, ensuring the regularity of $\chi$. Second, in the nonrelativistic regime, where $\mu \ll 1$, the anisotropic contribution becomes negligible, and the modified TOV equation smoothly reduces to the Newtonian hydrostatic equilibrium equation. Following earlier studies \cite{Doneva:2012rd,Folomeev:2018ioy,Silva:2014fca}, we consider values of the anisotropy parameter in the range $-2 \leq \tau_1 \leq 2$.

The second ansatz, proposed by Bowers and Liang \cite{Bowers:1974tgi,Becerra:2024xff}, is given by
\begin{equation}
	\chi = \tau_2\,(\rho + 3p_r)\,(\rho + p_r)\,
	\left(1-\frac{2m}{r}\right)^{-1} r^2 ,
\end{equation}
where $\tau_2$ characterizes the strength of anisotropy for this model. As in the previous case, we restrict the anisotropy parameter to the range $-2 \leq \tau_2 \leq 2$.

In constructing equilibrium configurations, we retain only physically acceptable and stable models. In particular, we impose the stability condition $\partial M/\partial \rho_c > 0$, where $\rho_c$ is the central density, and require the radial and tangential sound speeds ($v_{sr}^2 \equiv \partial p_r/\partial \rho$ and $v_{st}^2 \equiv \partial p_t/\partial \rho$, respectively) to satisfy $0 < v_{sr},\, v_{st} \leq 1$ throughout the stellar interior.

\begin{figure*}[t]
	\centering
	
	\begin{subfigure}[t]{0.49\textwidth}
		\centering
		\includegraphics[width=\linewidth]{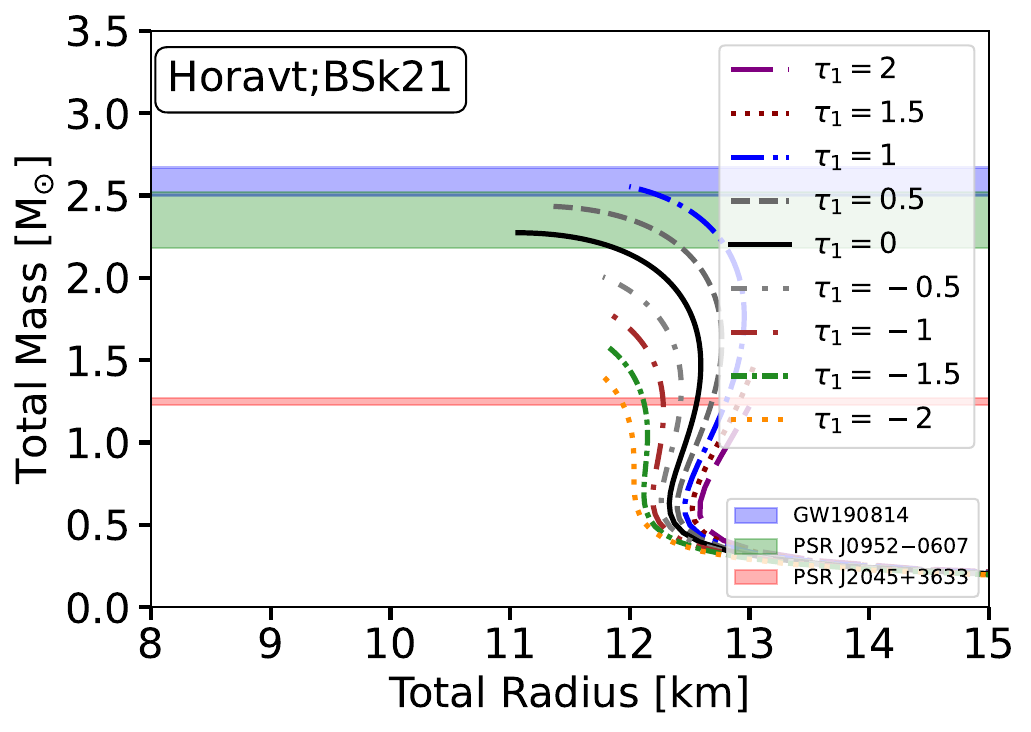}
		\caption{} \label{fig:mr_bsk21_hor}
	\end{subfigure}\hfill
	\begin{subfigure}[t]{0.49\textwidth}
		\centering
		\includegraphics[width=\linewidth]{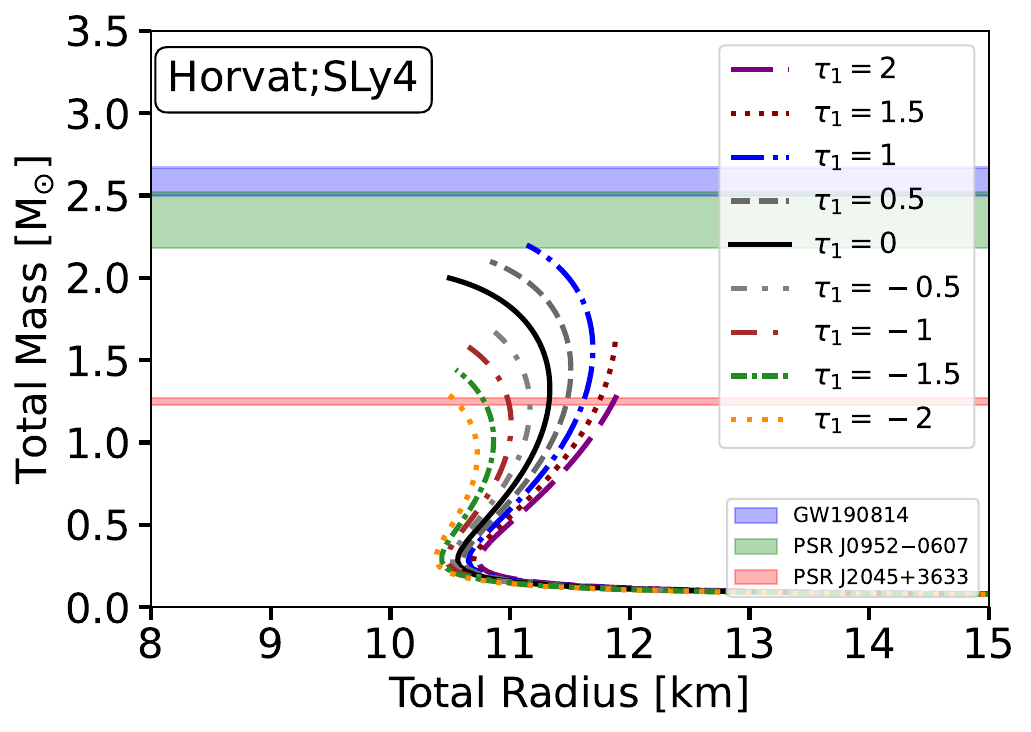}
		\caption{} \label{fig:mr_sly4_hor}
	\end{subfigure}
	
	\vspace{2mm}
	
	\begin{subfigure}[t]{0.49\textwidth}
		\centering
		\includegraphics[width=\linewidth]{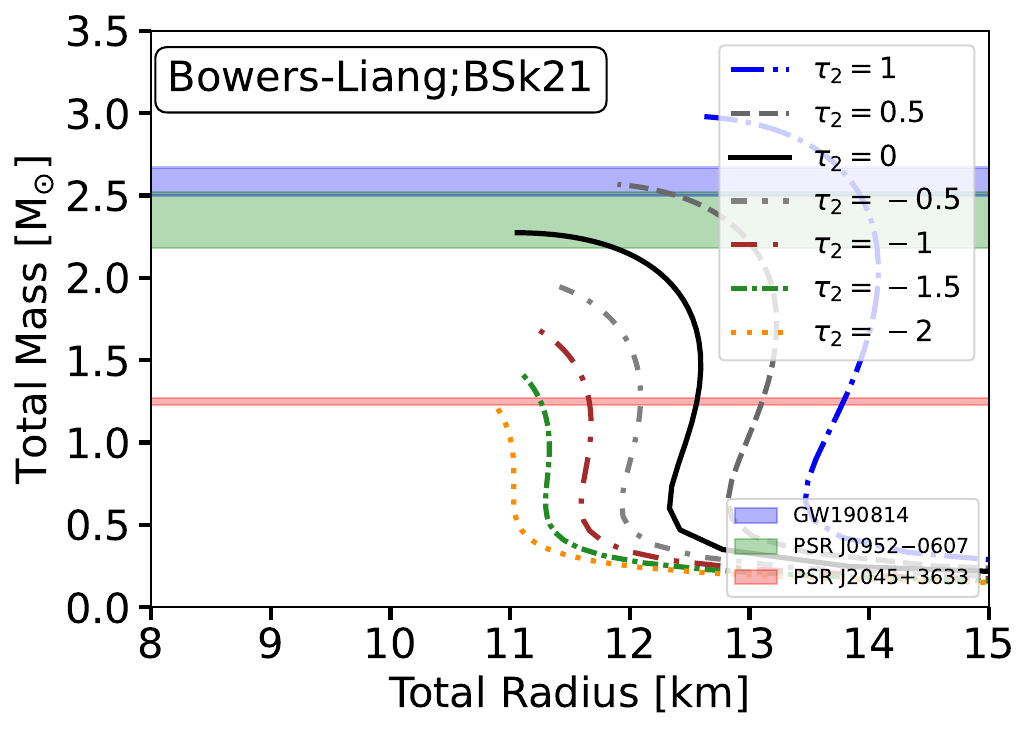}
		\caption{} \label{fig:mr_bsk21_bow}
	\end{subfigure}\hfill
	\begin{subfigure}[t]{0.49\textwidth}
		\centering
		\includegraphics[width=\linewidth]{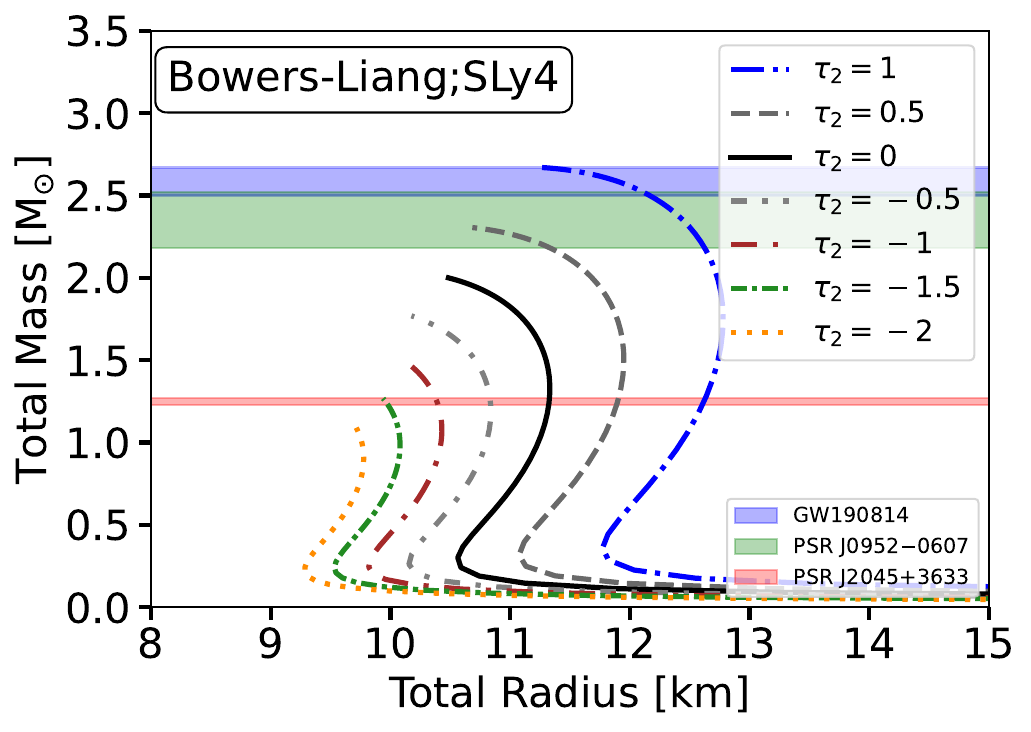}
		\caption{} \label{fig:mr_sly4_bow}
	\end{subfigure}
	
	\caption{Mass--radius relations for anisotropic neutron-star configurations constructed using the BSk21 and SLy4 equations of state, incorporating the two anisotropy prescriptions considered in this study. Panels~(a) and (b) correspond to the Horvat ansatz for BSk21 and SLy4, respectively, while panels~(c) and (d) display results for the Bowers--Liang ansatz. Curves are shown only for those values of the anisotropy parameter that yield stellar configurations satisfying the stability and physical acceptability criteria discussed in Sec.~\ref{sec:equilib}. Observational constraints are overlaid for comparison: the red band denotes the mass of PSR~J2045$+$3633~\cite{McKee:2020pzp}, a low-mass pulsar with $(1.252 \pm 0.021)\,M_\odot$; the green band indicates the high-mass pulsar PSR~J0952$-$0607~\cite{Romani:2022jhd}, with mass $(2.35 \pm 0.17)\,M_\odot$; and the blue band marks the secondary component of GW190814~\cite{LIGOScientific:2020zkf}, whose gravitational-wave-inferred mass lies within $2.50$--$2.67\,M_\odot$ (90\% credible interval).}

	\label{fig:MR_2x2}
\end{figure*}

In Fig.~\ref{fig:MR_2x2}, we show the mass--radius relations corresponding to all four possible combinations of the equations of state and anisotropy ansatzes discussed above. In each panel, only those neutron-star configurations that satisfy the physical acceptability and stability criteria outlined in Sec.~\ref{sec:equilib} are displayed.

For positive values of the anisotropy parameter ($\tau>0$, corresponding to $p_t>p_r$), the stellar mass initially increases with increasing $\tau$ and attains a maximum around $\tau=1$. For larger values of $\tau$, the mass decreases, as the corresponding configurations no longer satisfy the imposed physical conditions. In particular, beyond this threshold the square of the tangential sound speed, $v_{st}^2$, becomes negative, rendering such configurations unphysical. For the Bowers--Liang ansatz, stable neutron-star solutions are found only up to $\tau=1$ for both equations of state considered.

On the other hand, for negative anisotropy ($\tau<0$, corresponding to $p_t<p_r$), the total mass of the neutron star decreases monotonically with increasing magnitude of $\tau$. This qualitative behavior is common to both anisotropy ansatzes and both equations of state.

Quantitatively, for the Horvat ansatz the maximum neutron-star mass reaches $2.55\,M_\odot$ and $2.2\,M_\odot$ for the BSk21 and SLy4 equations of state, respectively. In comparison, the Bowers--Liang ansatz yields larger maximum masses, namely $2.98\,M_\odot$ for BSk21 and $2.67\,M_\odot$ for SLy4.

\section{Axial perturbations of anisotropic neutron stars}
\label{sec:axial_eq_der}

We consider linear perturbations of a static, spherically symmetric, nonrotating neutron star, described by a metric perturbation $h_{\alpha\beta}$ about the background spacetime. Owing to the spherical symmetry of the background, the perturbations can be decomposed into two independent sectors according to their behavior under parity transformations, namely the polar (even-parity) and axial (odd-parity) sectors. In the Regge--Wheeler gauge, the perturbation variables take a particularly simple form. The polar perturbations of anisotropic neutron stars, specifically $f$-modes, have been studied previously in Ref.~\cite{Mondal:2023wwo}. In the present work, we focus exclusively on the odd-parity (axial) sector.

The axial metric perturbation can be written as
\begin{equation}\label{axialmetric}
	h_{\mu\nu}^{(\mathrm{axial})} =
	\begin{pmatrix}
		0 & 0 & h_0\,S_\theta^{\ell m} & h_0\,S_\phi^{\ell m} \\
		0 & 0 & h_1\,S_\theta^{\ell m} & h_1\,S_\phi^{\ell m} \\
		h_0\,S_\theta^{\ell m} & h_1\,S_\theta^{\ell m} & 0 & 0 \\
		h_0\,S_\phi^{\ell m} & h_1\,S_\phi^{\ell m} & 0 & 0
	\end{pmatrix},
\end{equation}
where $h_0(t,r)$ and $h_1(t,r)$ are functions of the time coordinate $t$ and the radial coordinate $r$. The angular functions $S_\theta^{\ell m}$ and $S_\phi^{\ell m}$ are defined as
\begin{equation}
	S_\theta^{\ell m} = -\,\frac{1}{\sin\theta}\,\frac{\partial Y_{\ell m}}{\partial \phi},
	\qquad
	S_\phi^{\ell m} = \sin\theta\,\frac{\partial Y_{\ell m}}{\partial \theta},
\end{equation}
where $Y_{\ell m}(\theta,\phi)$ denote the scalar spherical harmonics. For brevity, in the following we omit the superscript ``axial'' and denote the metric perturbation simply by $h_{\mu\nu}$.

To describe the fluid displacement associated with axial perturbations, we introduce the displacement vector
\begin{equation}\label{xiexp}
	\xi^\mu = \Lambda(t,r)\,(0,\,0,\,S_\theta^{\ell m},\,S_\phi^{\ell m})^T ,
\end{equation}
where $\Lambda(t,r)$ is a scalar function of $t$ and $r$, and the angular functions $S_\theta^{\ell m}$ and $S_\phi^{\ell m}$ are defined above. Following the formalism described in Refs.~\cite{Andersson:2006nr,Mondal:2023wwo}, the perturbation of the four-velocity $u^\mu$ can be expressed as
\begin{equation}\label{pertvel}
	\delta u^\mu = (\delta^{\mu}{}_{\rho} + u^\mu u_\rho)
	\left( u^\sigma \nabla_\sigma \xi^\rho - \xi^\sigma \nabla_\sigma u^\rho \right)
	+ \frac{1}{2} u^\mu u^\sigma u^\rho h_{\sigma\rho}.
\end{equation}

Using the background four-velocity $u^\alpha$ given in Eq.~(\ref{uands}), the displacement vector $\xi^\alpha$ from Eq.~(\ref{xiexp}), and the axial metric perturbation $h_{\alpha\beta}$ from Eq.~(\ref{axialmetric}), the explicit form of the perturbed four-velocity is obtained as
\begin{equation}
	\delta u^\alpha = e^{-\nu/2}\,\dot{\Lambda}
	\left(0,\,0,\,r^2 S_\theta^{\ell m},\,\frac{S_\phi^{\ell m}}{r^2 \sin^2\theta}\right)^T ,
\end{equation}
where an overdot denotes differentiation with respect to the time coordinate $t$.

From the normalization conditions of the spacelike unit vector $s^\alpha$ and the four-velocity $u^\alpha$, given in Eq.~(\ref{usnorm}), we have
\begin{equation}\label{normpert}
	\delta(s_\alpha s^\alpha) = 0, \qquad
	\delta(s_\alpha u^\alpha) = 0 .
\end{equation}
Imposing these conditions and using the background expressions for $u^\alpha$ and $s^\alpha$ from Eq.~(\ref{uands}), one finds
\begin{equation}
	\delta s^\alpha = (0,\,0,\,0,\,0)^T .
\end{equation}
It can be readily verified that this result is consistent with the perturbed normalization conditions given in Eq.~(\ref{normpert}).

Having established the necessary background quantities, we now proceed to derive the governing equations for axial perturbations of neutron stars by linearizing the Einstein equations about the static, spherically symmetric background spacetime. The linearized Einstein tensor can be expressed as \cite{Kojima:1992ie}
\begin{equation}\label{perteinst}
	\begin{split}
		\delta G_{\mu\nu} &= -\frac{1}{2}\Big[ \,
		\nabla^\alpha \nabla_\alpha h_{\mu\nu}
		- \left(\nabla_\nu f_\mu + \nabla_\mu f_\nu\right)
		+ 2\,{\mathcal R^\alpha}_{\ \mu}{}^\beta{}_{\ \nu}\,h_{\alpha\beta} \\
		&+ \nabla_\nu \nabla_\mu {h^\alpha}_{\alpha}
		- \left({\mathcal R^\alpha}_{\ \nu}\,h_{\mu\alpha}
		+ {\mathcal R^\alpha}_{\ \mu}\,h_{\nu\alpha}\right)
		+ g^{(B)}_{\mu\nu}\left(\nabla^\alpha f_\alpha \right. \\
		&\left.\quad + \nabla^\beta \nabla_\beta {h^\alpha}_{\alpha}\right)
		+ \mathcal R\, h_{\mu\nu}
		- g^{(B)}_{\mu\nu}\,{\mathcal R}^{\alpha\beta} h_{\alpha\beta}
		\Big] .
	\end{split}
\end{equation}

where $f_\nu = \nabla^\beta h_{\nu\beta}$, $\mathcal{R}_{\alpha\beta\gamma\delta}$, $\mathcal{R}_{\alpha\beta}$, and $\mathcal{R}$ denote the Riemann tensor, Ricci tensor, and Ricci scalar of the background spacetime, respectively, and $\nabla_\sigma$ represents the covariant derivative associated with the background metric $g^{(B)}_{\mu\nu}$. The linearized energy--momentum tensor for an anisotropic fluid can be written as
\begin{equation}
	\begin{split}
		\delta T_{\mu \nu} = &(\delta \rho + \delta p_t)\, u_\mu u_\nu
		+ (\rho + p_t)\,(u_\mu \delta u_\nu + u_\nu \delta u_\mu)
		+ g^{(B)}_{\mu \nu}\,\delta p_t \\
		&+ p_t\,\delta g^{(B)}_{\mu\nu}
		+ (p_r - p_t)\,(s_\mu \delta s_\nu + s_\nu \delta s_\mu) .
	\end{split}
\end{equation}
Substituting the explicit expressions for the perturbed quantities into the above relations and inserting the result into the linearized Einstein equations, we obtain the equations governing axial perturbations. In particular, from the $(1,3)$ component of the field equations, $\delta G_{13} = 8\pi\,\delta T_{13}$, we find
\begin{equation}
	\frac{\partial^2 h_1}{\partial t^2}
	+ \left(\frac{2}{r} - \frac{\partial}{\partial r}\right)
	\frac{\partial h_0}{\partial t}
	+ e^{\nu}\,\frac{(\ell -1)(\ell +2)}{r^2}\,h_1
	= 0 .
\end{equation}
Similarly, the $(2,3)$ component of the linearized Einstein equations, $\delta G_{23} = 8\pi\,\delta T_{23}$, yields
\begin{equation}
	\frac{\partial h_0}{\partial t}
	- e^{(\nu - \lambda)/2}\,
	\frac{\partial}{\partial r}
	\left(e^{(\nu - \lambda)/2} h_1 \right)
	= 0 .
\end{equation}
Combining the above two equations, one arrives at a single wave equation of the form
\begin{equation}
	\begin{split}
		\frac{\partial^2 X}{\partial t^2}
		&- e^{(\nu-\lambda)/2}\,
		\frac{\partial}{\partial r}
		\left(
		e^{(\nu-\lambda)/2}
		\frac{\partial X}{\partial r}
		\right) \\
		&+ e^{\nu}
		\left[
		\frac{\ell(\ell+1)}{r^2}
		- \frac{3}{r^2}\left(1-e^{-\lambda}\right)
		+ 4\pi\left(\rho - p_r\right)
		\right] X
		= 0 ,
	\end{split}
\end{equation}
where we have introduced the master variable
\begin{equation}
	X = \frac{e^{(\nu - \lambda)/2}}{r}\,h_1 .
\end{equation}

Assuming a harmonic time dependence of the form
\begin{equation}
	X(r,t) = \bar{X}(r)\,e^{i\omega t},
\end{equation}
the above equation can be recast as
\begin{equation}\label{eq:axial}
	\frac{d^2 \bar{X}}{d r_*^2}
	+ \left[
	\omega^2
	- \frac{e^{\nu}}{r^3}
	\left(
	\ell(\ell+1)r
	- 6m
	+ 4\pi r^3 (\rho - p_r)
	\right)
	\right]\bar{X}
	= 0 ,
\end{equation}
where $r_*$ denotes the tortoise coordinate, defined through
\begin{equation}
	\frac{d r_*}{d r} = e^{(\lambda - \nu)/2}.
\end{equation}

To extract the axial $w$-mode spectrum, Eq.~(\ref{eq:axial}) must be integrated from the stellar center to spatial infinity. Regularity of the perturbation at the stellar center requires a power-series expansion of the master variable. Near the center, the solution takes the form
\begin{equation}
	\begin{split}
		\bar{X} = r^{\ell + 1}
		+ \frac{1}{2 \ell + 3}
		\Bigg[
		4 \pi (\ell + 2)
		\left(
		\frac{2 \ell - 1}{3}\,\rho_0
		- p_{r0}
		\right) \\
		\qquad
		- \omega^2 e^{-\nu_0}
		\Bigg]
		r^{\ell + 3}~,
	\end{split}
\end{equation}
where $\rho_0$, $p_{r0}$, and $\nu_0$ denote the leading-order coefficients in the near-center expansions of the energy density $\rho$, radial pressure $p_r$, and metric function $\nu$, respectively. The above series solution is valid only in the immediate vicinity of the stellar center. In practice, we initialize the integration at a small but finite radius, typically of order $r \sim R/10^6$, where $R$ is the stellar radius, and use this solution as the inner boundary condition. From this radius, Eq.~(\ref{eq:axial}) is integrated numerically up to the stellar surface.

Outside the star, where $\rho = p_r = 0$, the axial perturbation equation reduces to its vacuum form,
\begin{equation}\label{eq:axial_out}
	\frac{d^2 \bar{X}}{d r_*^2}
	+ \left[
	\omega^2
	- \frac{(r - 2M)}{r^4}
	\left(
	\ell(\ell+1)r - 6M
	\right)
	\right]\bar{X}
	= 0 ,
\end{equation}
where $M$ is the total mass of the star.

In principle, Eq.~(\ref{eq:axial_out}) must be integrated from the stellar surface to spatial infinity, with a purely outgoing-wave boundary condition imposed at infinity. However, for axial $w$-modes the imaginary part of the frequency is typically large, which renders a direct numerical implementation of this boundary condition unstable. To overcome this difficulty, we employ the continued-fraction (Leaver) method, which has been discussed extensively in Refs.~\cite{Leaver:1985,Leins:1993zz}. Here we will describe some crucial steps of this method.

\begin{figure*}[t]
	\centering
	
	\begin{subfigure}[t]{0.49\textwidth}
		\centering
		\includegraphics[width=\linewidth]{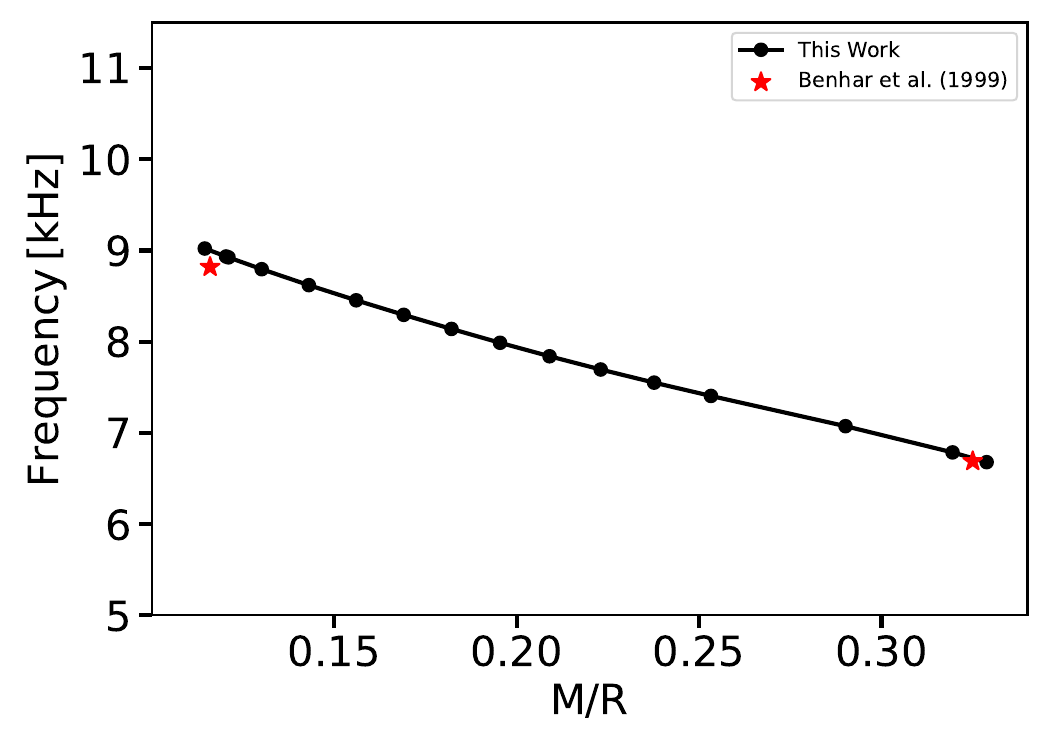}
		\caption{} \label{fig:freq_compactness_APR}
	\end{subfigure}\hfill
	\begin{subfigure}[t]{0.49\textwidth}
		\centering
		\includegraphics[width=\linewidth]{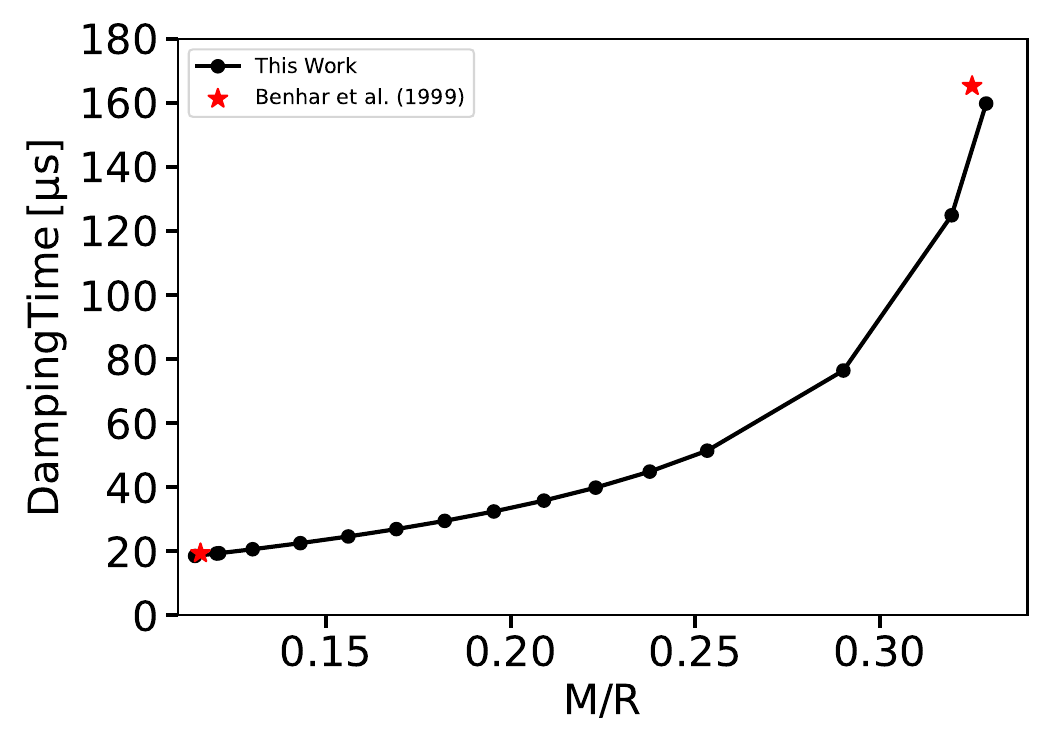}
		\caption{} \label{fig:dt_compactness_APR}
	\end{subfigure}
	
	\caption{Frequency-compactness and damping 
			time-compactness relations of the fundamental axial 
			$w$-mode computed using the APR2 equation of state 
			\cite{Akmal:1998} (including special relativistic corrections) in the isotropic limit. Panel (a) 
			shows the mode frequency as a function of 
			the stellar compactness $\mathrm{M/R}$, while panel (b) shows 
			the corresponding damping time. In both 
			panels, filled circles connected by a solid line 
			represent our numerical results, and filled stars denote 
			the benchmark data points from Table~1 of 
			\citet{Benhar:1998au} for the same equation of state. 
			The close agreement between the two sets of results 
			validates our numerical implementation.}
	\label{fig:validation}
\end{figure*}

In this approach, Eq.~(\ref{eq:axial_out}) is first integrated numerically from the stellar surface to a matching radius $R_f$, chosen such that $R < R_f < 2R$ \cite{Sotani:2001bb}. The exterior solution is then expressed as a Leaver series of the form
\begin{equation}
	\bar{X}(r)
	= \left(\frac{r}{2M} - 1\right)^{-2 i M \omega}
	e^{-i \omega r}
	\sum_{n=0}^{\infty} a_n
	\left(1 - \frac{R_f}{r}\right)^n .
\end{equation}
Substituting this expansion into Eq.~(\ref{eq:axial_out}) leads to a four-term recurrence relation among the coefficients $a_n$, which is given by
\begin{equation}
	\alpha_n a_{n+1} + \beta_n a_{n} + \gamma_n a_{n-1} + \delta_n a_{n-2} = 0~,
\end{equation}
where
\begin{subequations}
	\begin{align}
		\alpha_n &= \left(1 - \frac{2M}{R_f}\right) n(n+1) , \\
		\beta_n  &= -2 \left[ i \omega R_f
		+ \left(1 - \frac{3M}{R_f}\right) n \right] n , \\
		\gamma_n &= \left(1 - \frac{6M}{R_f}\right) n(n-1)
		+ \frac{6M}{R_f}
		- \ell(\ell +1) , \\
		\delta_n &= \frac{2M}{R_f} (n-3)(n+1) .
	\end{align}
\end{subequations}
The coefficients $a_0$ and $a_1$ are determined by the continuity of $\bar{X}$ and $\frac{d \bar{X}}{d r}$ at $r = R_f$, that is
\begin{subequations}
	\begin{align}
		a_0 &= \frac{\bar{X}(R_f)}{\mathfrak{A}(R_f)} ,\\
		a_1
		&= \frac{R_f}{\mathfrak{A}(R_f)}
		\left[
		\left.\frac{d\bar{X}}{dr}\right|_{R_f}
		+ \frac{i\omega R_f}{R_f - 2M}\,\bar{X}(R_f)
		\right] ,
	\end{align}
	
\end{subequations}
where 
\begin{equation}
	\mathfrak{A}(r)
	= \left(\frac{r}{2M} - 1\right)^{-2 i M \omega}
	\, e^{-i \omega r} .
\end{equation}
The four-term recurrence relation can be reduced to a three-term one by defining
\begin{equation}
	\tilde{\alpha}_1 = \alpha_1, \qquad
	\tilde{\beta}_1 = \beta_1, \qquad
	\tilde{\gamma}_1 = \gamma_1 .
\end{equation}
For $n \geq 2$, the modified coefficients are given by
\begin{equation}
	\tilde{\alpha}_n = \alpha_n ,
\end{equation}
\begin{equation}
	\tilde{\beta}_n = \beta_n
	- \frac{\tilde{\alpha}_{n-1}\,\delta_n}{\tilde{\gamma}_{n-1}} ,
\end{equation}
\begin{equation}
	\tilde{\gamma}_n = \gamma_n
	- \frac{\tilde{\beta}_{n-1}\,\delta_n}{\tilde{\gamma}_{n-1}} ,
\end{equation}
together with
\begin{equation}
	\tilde{\delta}_n = 0 .
\end{equation}
With these definitions, the original four-term recurrence relation is transformed into the three-term recurrence relation
\begin{equation}
	\tilde{\alpha}_n a_{n+1}
	+ \tilde{\beta}_n a_n
	+ \tilde{\gamma}_n a_{n-1}
	= 0 .
\end{equation}
The three-term recurrence relation leads to the following continued-fraction condition,
\begin{equation}
	\frac{a_1}{a_0}
	= \frac{-\tilde{\gamma}_1}
	{\tilde{\beta}_1
		- \dfrac{\tilde{\alpha}_1 \tilde{\gamma}_2}
		{\tilde{\beta}_2
			- \dfrac{\tilde{\alpha}_2 \tilde{\gamma}_3}{\tilde{\beta}_3 - \cdots}}} .
\end{equation}
This expression can be rewritten in the compact form
\begin{equation}
	0 = \tilde{\beta}_0
	- \frac{\tilde{\alpha}_0 \tilde{\gamma}_1}
	{\tilde{\beta}_1
		- \dfrac{\tilde{\alpha}_1 \tilde{\gamma}_2}
		{\tilde{\beta}_2
			- \dfrac{\tilde{\alpha}_2 \tilde{\gamma}_3}{\tilde{\beta}_3 - \cdots}}}
	\equiv f(\omega) ,
\end{equation}
where $\tilde{\beta}_0 = a_1/a_0$ and $\tilde{\alpha}_0 = -1$. The complex quasinormal-mode frequency $\omega$ is determined by solving the equation $f(\omega)=0$. Although the above formalism is valid for arbitrary values of the multipole index $\ell$, in the present work we restrict our analysis to $\ell = 2$, as this mode dominates gravitational-wave emission.

In practice, the function $f(\omega)$ is evaluated over a region of the complex $\omega$ plane by assigning trial complex values to $\omega$ and computing the corresponding continued fraction. The roots of $f(\omega)$ are then located using standard root-finding techniques, and the values of $\omega$ for which $f(\omega)=0$ define the axial $w$-mode spectrum.

The real part of the complex frequency, $\mathrm{Re}(\omega)$, corresponds to the oscillation frequency of the mode, while the inverse of the imaginary part, $1/\mathrm{Im}(\omega)$, gives the associated damping time. Further details of the numerical implementation of this procedure can be found in Ref.~\cite{Sotani:2001bb}.
  
\section{Numerical results for axial w-modes}
\label{sec:numerical_results}

\begin{figure*}[t]
	\centering
	
	\begin{subfigure}[t]{0.49\textwidth}
		\centering
		\includegraphics[width=\linewidth]{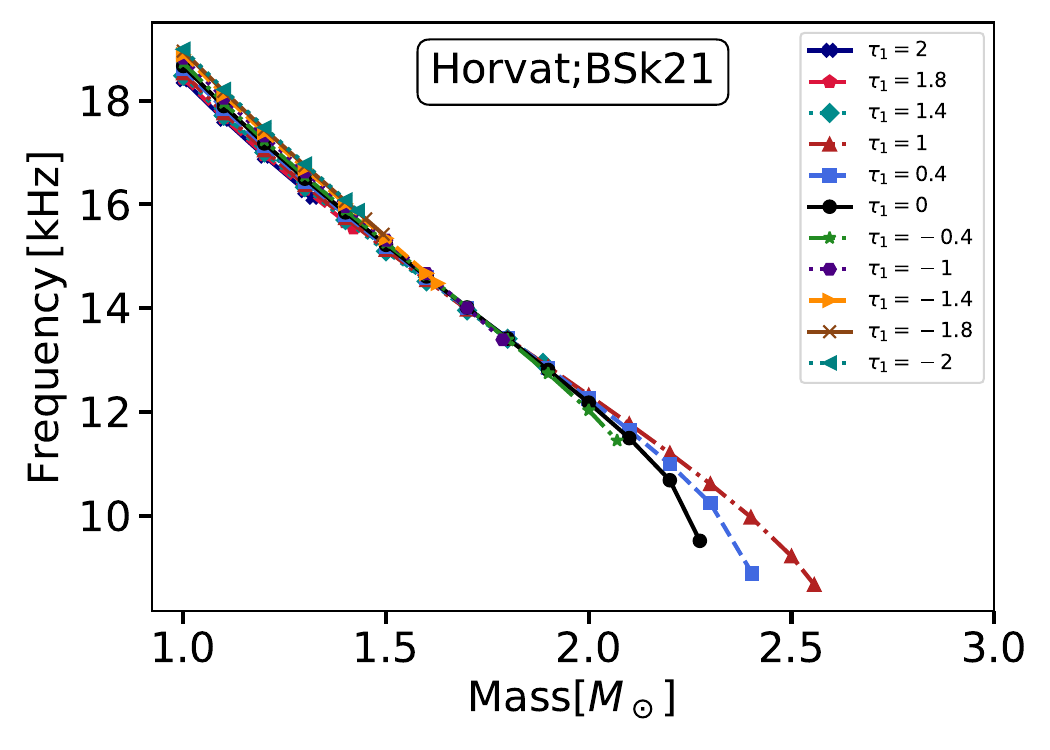}
		\caption{} \label{fig:freq_mass_bsk21_hor}
	\end{subfigure}\hfill
	\begin{subfigure}[t]{0.49\textwidth}
		\centering
		\includegraphics[width=\linewidth]{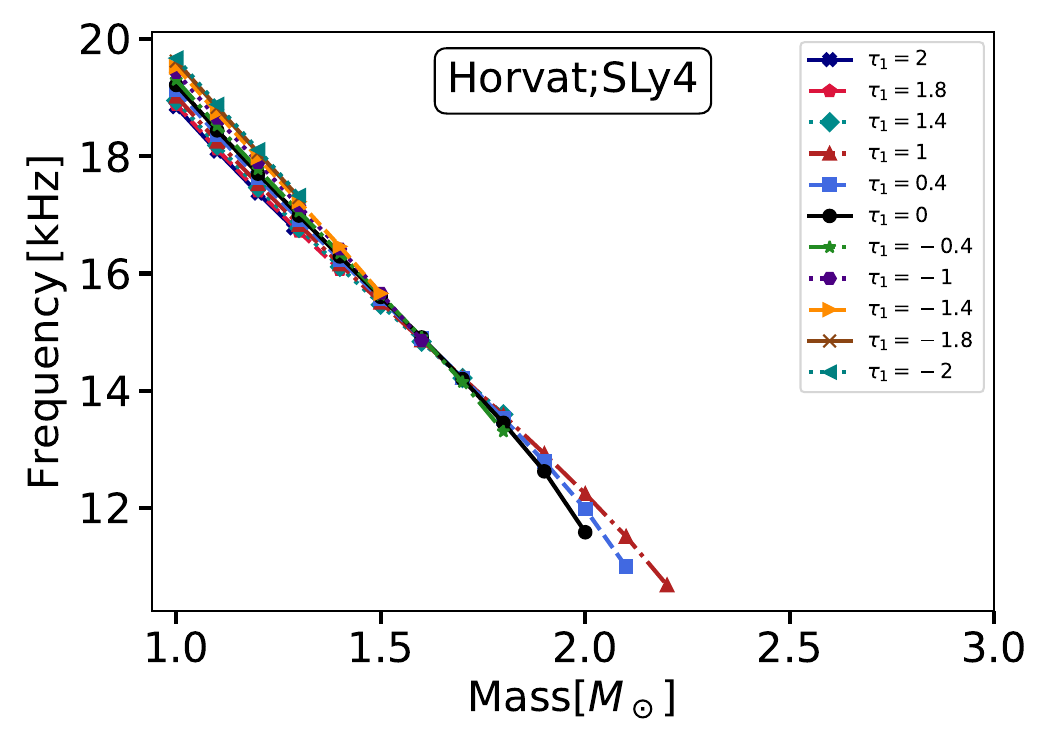}
		\caption{} \label{fig:freq_mass_sly4_hor}
	\end{subfigure}
	
	\vspace{2mm}
	
	\begin{subfigure}[t]{0.49\textwidth}
		\centering
		\includegraphics[width=\linewidth]{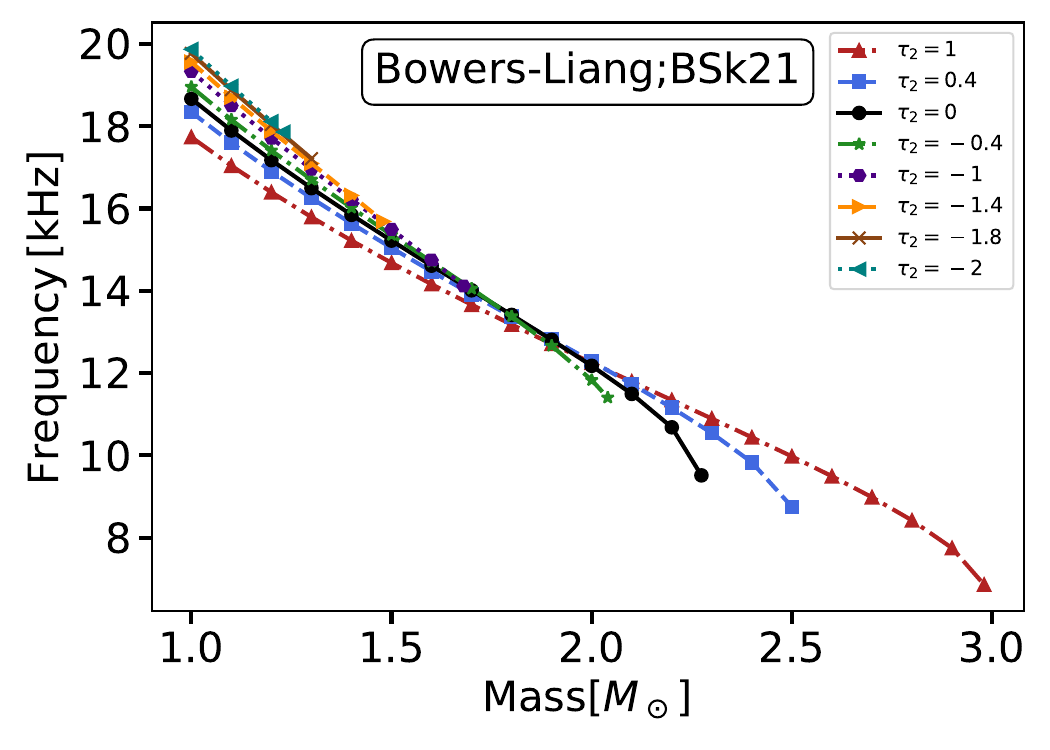}
		\caption{} \label{fig:freq_mass_bsk21_bow}
	\end{subfigure}\hfill
	\begin{subfigure}[t]{0.49\textwidth}
		\centering
		\includegraphics[width=\linewidth]{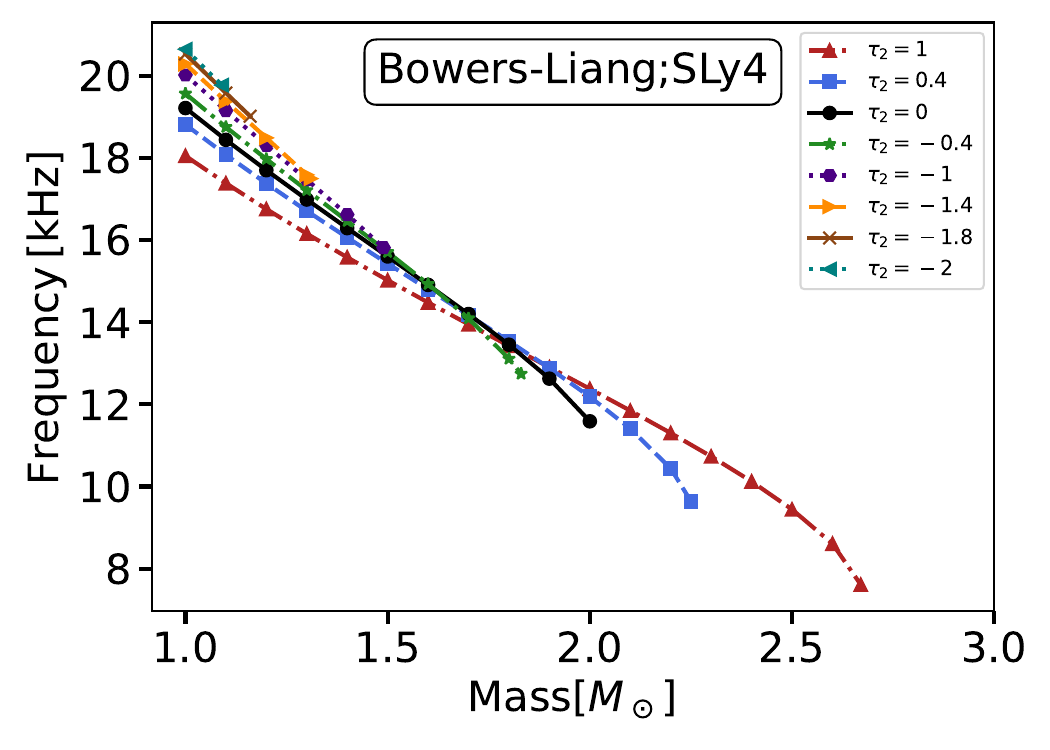}
		\caption{} \label{fig:freq_mass_sly4_bow}
	\end{subfigure}
	
	\caption{Frequency-mass relations of axial $w$-modes for anisotropic neutron-star configurations constructed using the BSk21 and SLy4 equations of state. Panels (a) and (b) correspond to the Horvat ansatz for BSk21 and SLy4, respectively, while panels (c) and (d) show the corresponding results obtained using the Bowers--Liang ansatz. Curves are shown only for those values of the anisotropy parameter for which the stellar models satisfy the stability and physical acceptability conditions adopted in Sec.~\ref{sec:equilib}.}
	\label{fig:freq-mass_2x2}
\end{figure*}
In this section, we present the results of our investigation of the first (lowest) axial $w$-mode oscillations of anisotropic neutron stars for the equations of state and anisotropy ansatzes described in  Sec.~\ref{sec:equilib}. We denote the oscillation frequency by $\mathcal{F}$ and the damping time by $\mathfrak{T}$, where $\mathcal{F} = \mathrm{Re}(\omega)/(2\pi)$ and $\mathfrak{T} = [\mathrm{Im}(\omega)]^{-1}$. Here, $\omega$ is the complex frequency appearing in the oscillation equations derived above, with $\mathrm{Re}(\omega)$ and $\mathrm{Im}(\omega)$ representing its real and imaginary parts, respectively.

Before presenting the main results, we validate our numerical implementation by comparing the 
axial $w$-mode frequency and damping time computed in the 
isotropic limit with the benchmark results of \citet{Benhar:1998au}. Specifically, we use the APR2 
equation of state \cite{Akmal:1998} (including special relativistic corrections) with vanishing 
anisotropy and compare our results against the two data 
points tabulated in Table~1 of \citet{Benhar:1998au} for the same equation of state. As shown in Fig.~\ref{fig:validation}, our results are in good 
agreement with those of \citet{Benhar:1998au} for both 
the frequency and the damping time, confirming the 
validity of our calculation. Furthermore, the overall 
trends of the frequency and damping time as functions of 
the compactness, shown in Fig.~\ref{fig:validation}, are 
consistent with those displayed in Figs.~5 and~6 of 
\citet{Benhar:1998au}.

Since the gravitational mass is one of the most directly accessible macroscopic observables, in Fig.~\ref{fig:freq-mass_2x2} we present the axial $w$-mode frequency as a function of the stellar mass for various values of the anisotropy strength, constructed using the BSk21 and SLy4 equations of state and both anisotropy ansatzes considered in this work. In each panel, the mode frequency decreases monotonically along the stable branch as the stellar mass increases.

\begin{figure*}[t]
	\centering
	
	\begin{subfigure}[t]{0.49\textwidth}
		\centering
		\includegraphics[width=\linewidth]{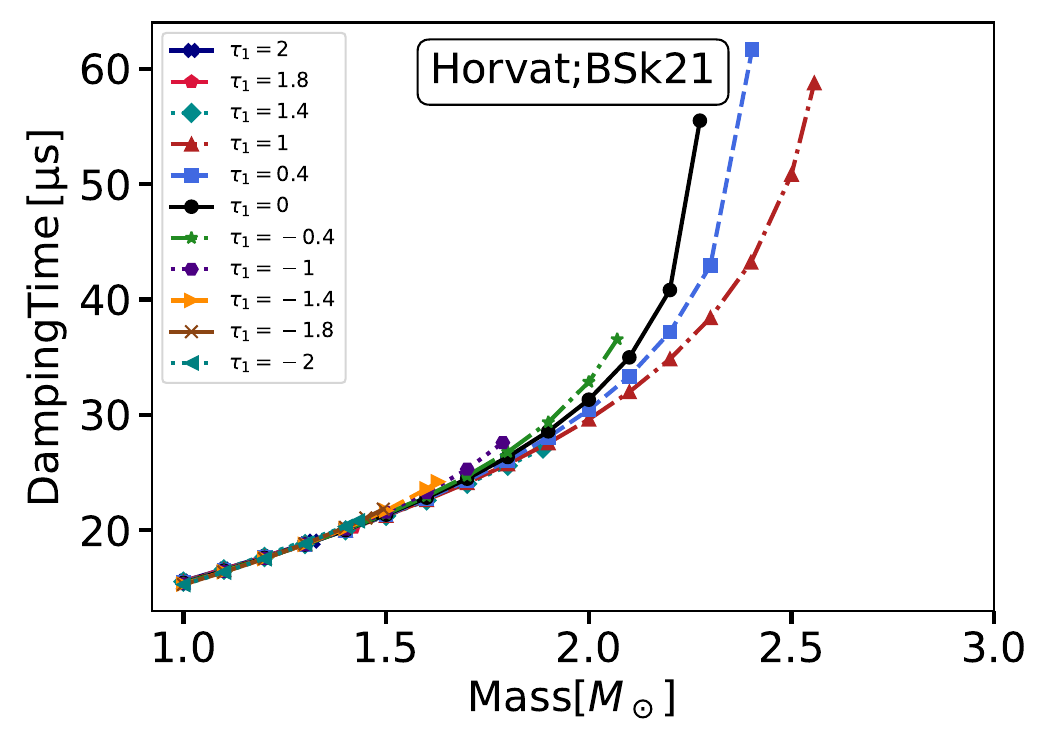}
		\caption{} \label{fig:dt_mass_bsk21_hor}
	\end{subfigure}\hfill
	\begin{subfigure}[t]{0.49\textwidth}
		\centering
		\includegraphics[width=\linewidth]{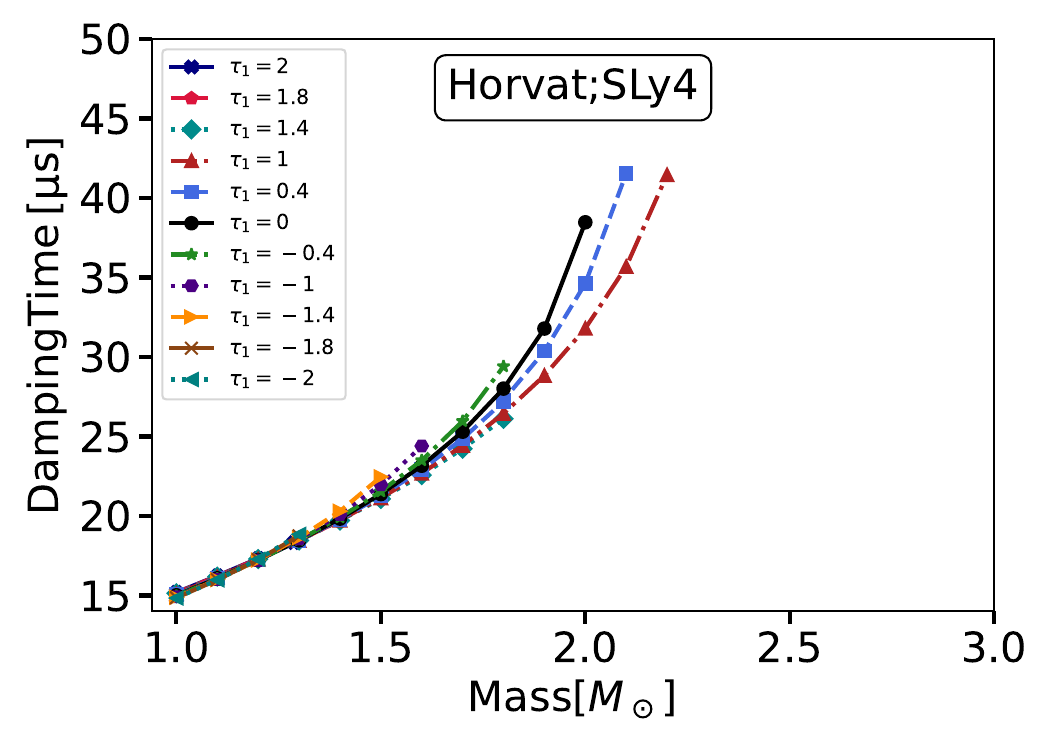}
		\caption{} \label{fig:dt_mass_sly4_hor}
	\end{subfigure}
	
	\vspace{2mm}
	
	\begin{subfigure}[t]{0.49\textwidth}
		\centering
		\includegraphics[width=\linewidth]{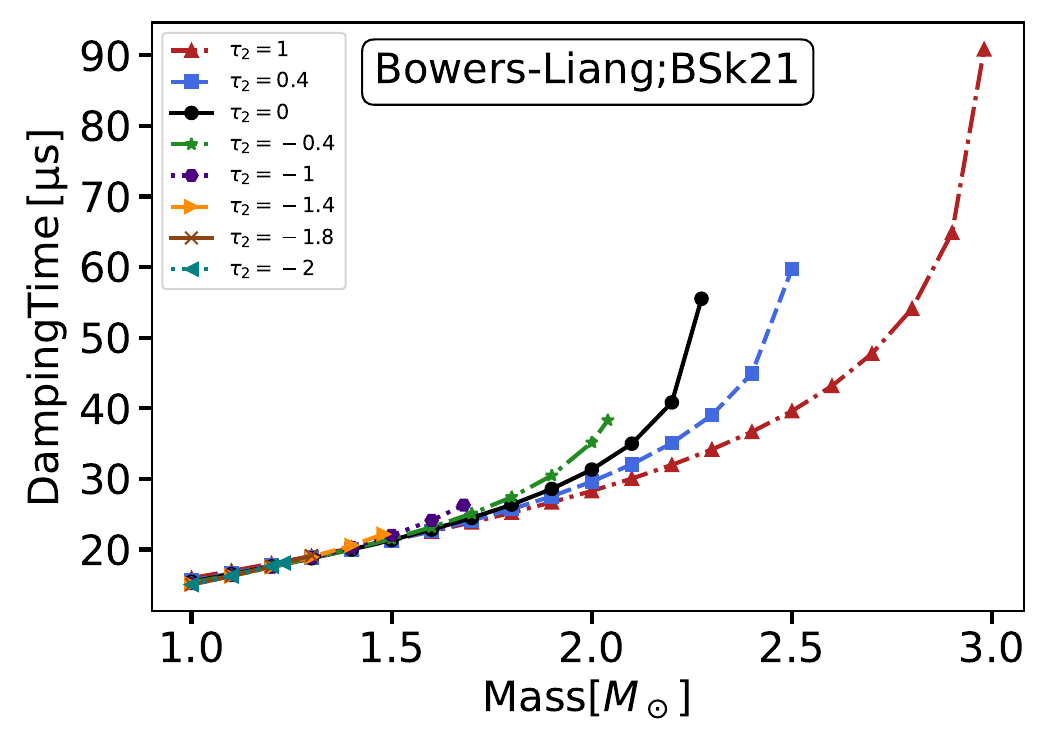}
		\caption{} \label{fig:dt_mass_bsk21_bow}
	\end{subfigure}\hfill
	\begin{subfigure}[t]{0.49\textwidth}
		\centering
		\includegraphics[width=\linewidth]{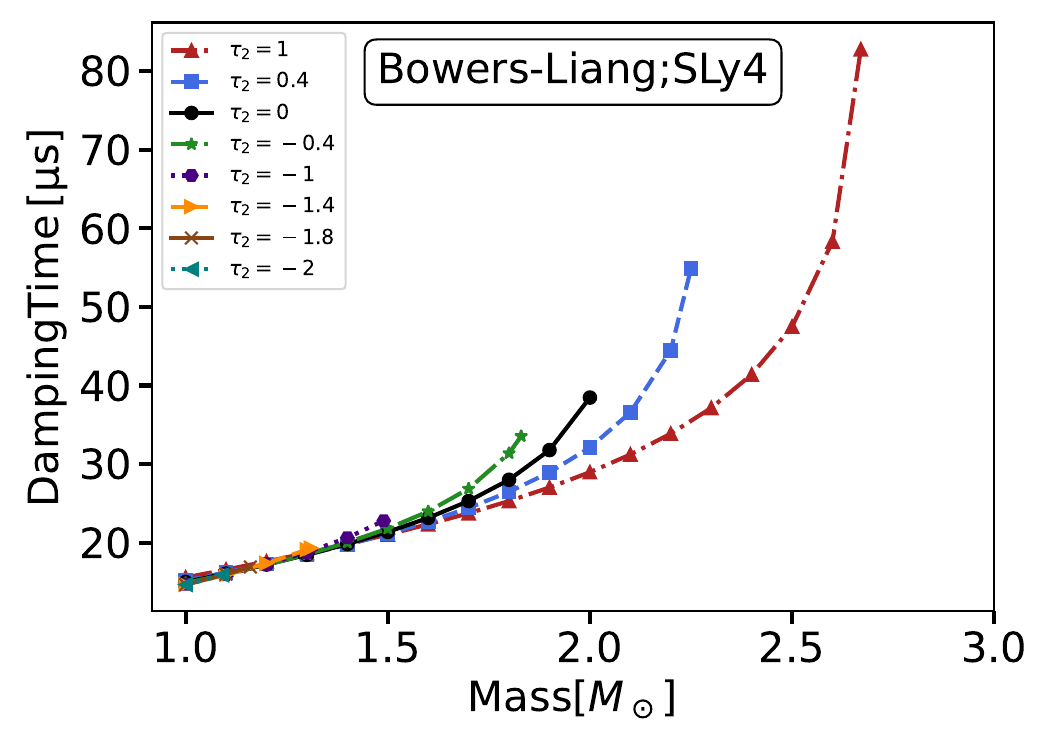}
		\caption{} \label{fig:dt_mass_sly4_bow}
	\end{subfigure}
	
	\caption{Damping time-mass relations of axial $w$-modes for anisotropic neutron-star configurations constructed using the BSk21 and SLy4 equations of state. Panels (a) and (b) correspond to the Horvat ansatz for BSk21 and SLy4, respectively, while panels (c) and (d) show the corresponding results obtained using the Bowers--Liang ansatz. Curves are shown only for those values of the anisotropy parameter for which the stellar models satisfy the stability and physical acceptability conditions adopted in Sec.~\ref{sec:equilib}.}
	\label{fig:dt-mass_2x2}
\end{figure*}

The dependence on anisotropy exhibits a clear mass-dependent trend. For relatively low stellar masses, close to $M \sim 1\,M_\odot$, configurations with $p_r>p_t$ (negative anisotropy, $\tau_i<0$) yield larger axial-mode frequencies at a fixed mass than those with $p_t>p_r$ (positive anisotropy, $\tau_i>0$). As the stellar mass increases, this ordering is reversed: near the upper end of the stable branch (typically around $M \sim 2\,M_\odot$ for the sequences shown), configurations with $p_t>p_r$ (positive anisotropy) exhibit higher frequencies than their negative-anisotropy counterparts at the same mass. This change in ordering is observed for both equations of state and for both anisotropy ansatzes, and is most pronounced toward the termination of each sequence, where the effects of anisotropy become more significant.

For the Horvat ansatz [panels (a) and (b)], the low-mass curves are nearly degenerate, so anisotropy-induced differences are less pronounced in this regime and become appreciable only at larger stellar masses. In contrast, the Bowers--Liang ansatz [panels (c) and (d)] exhibits a clearer spread across anisotropy strengths over a broader mass interval. In all cases, the endpoint of each curve is determined by the subset of equilibrium configurations that satisfy the stability and physical acceptability conditions adopted in Sec.~\ref{sec:equilib}, and therefore different anisotropy strengths terminate at different maximum masses.

\begin{figure*}[t]
	\centering
	
	\begin{subfigure}[t]{0.49\textwidth}
		\centering
		\includegraphics[width=\linewidth]{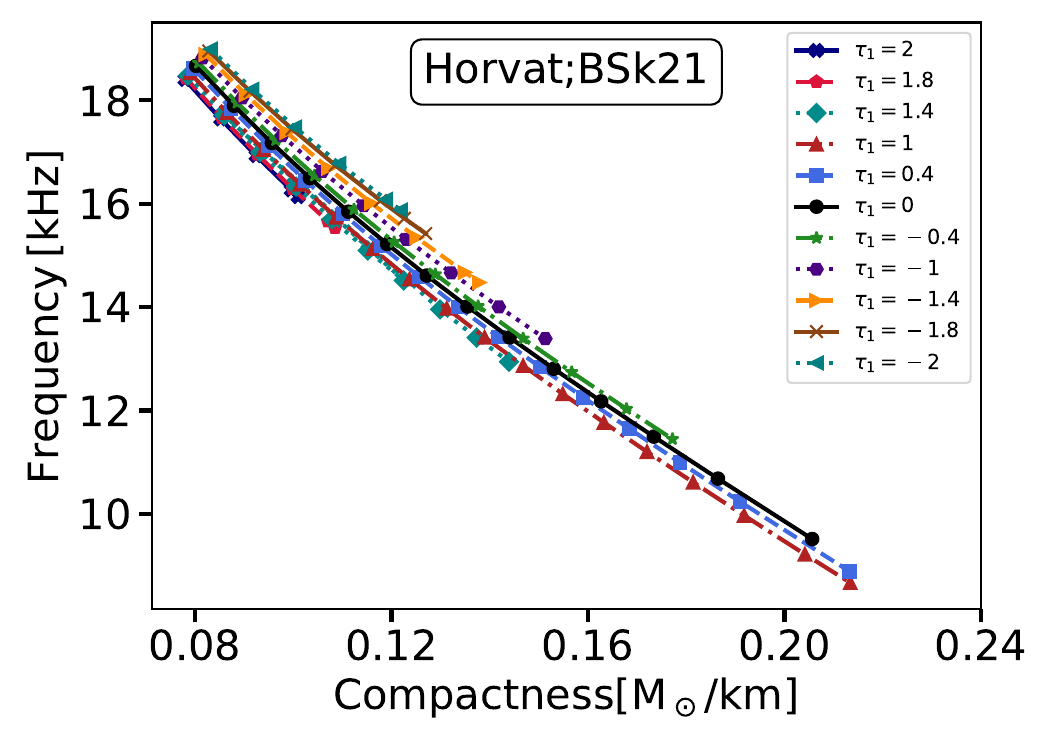}
		\caption{} \label{fig:freq_compactness_bsk21_hor}
	\end{subfigure}\hfill
	\begin{subfigure}[t]{0.49\textwidth}
		\centering
		\includegraphics[width=\linewidth]{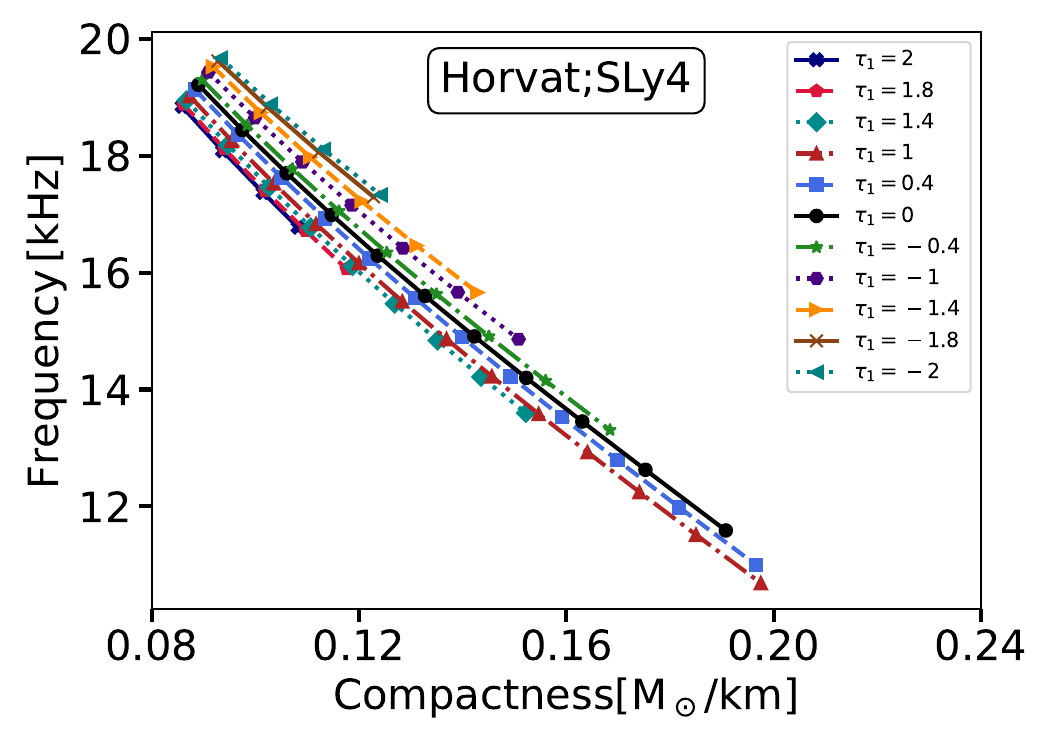}
		\caption{} \label{fig:freq_compactness_sly4_hor}
	\end{subfigure}
	
	\vspace{2mm}
	
	\begin{subfigure}[t]{0.49\textwidth}
		\centering
		\includegraphics[width=\linewidth]{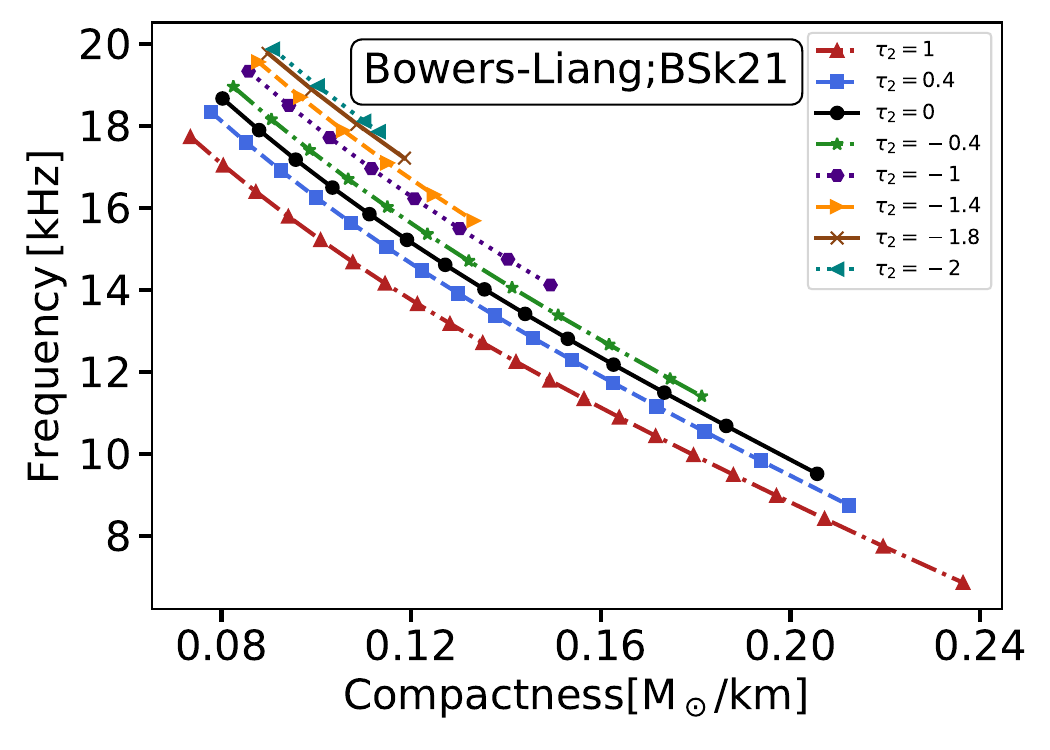}
		\caption{} \label{fig:freq_compactness_bsk21_bow}
	\end{subfigure}\hfill
	\begin{subfigure}[t]{0.49\textwidth}
		\centering
		\includegraphics[width=\linewidth]{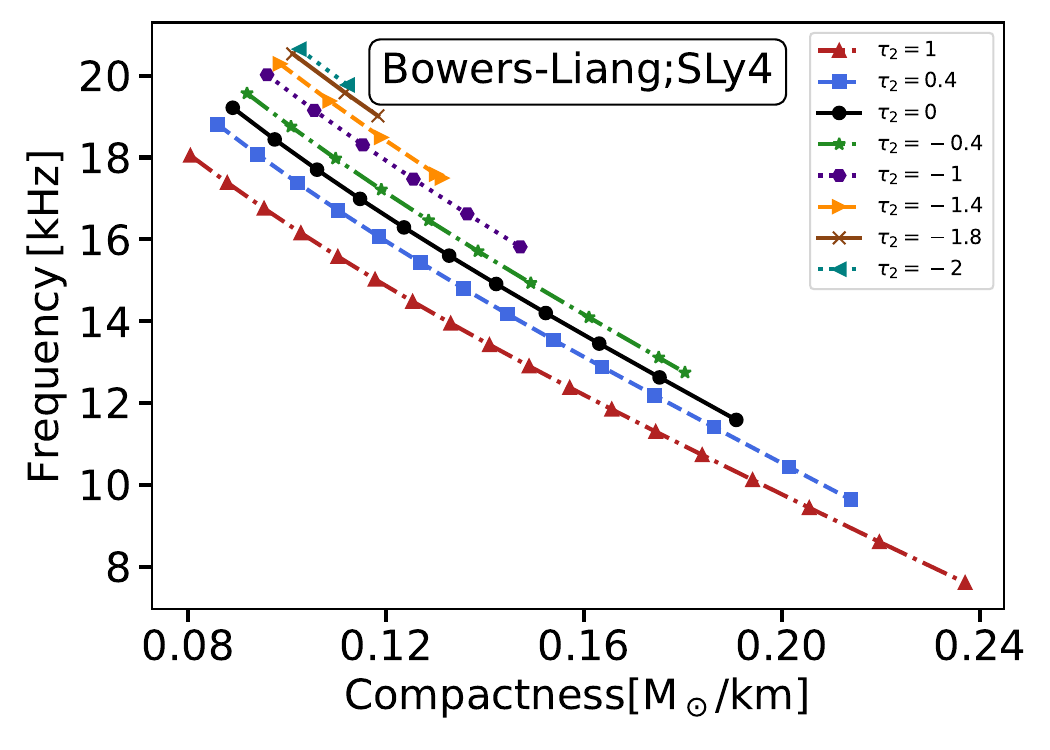}
		\caption{} \label{fig:freq_compactness_sly4_bow}
	\end{subfigure}
	
	\caption{Frequency-compactness relations of axial $w$-modes for anisotropic neutron-star models constructed with the BSk21 and SLy4 equations of state. The upper panels [(a), (b)] show results obtained using the Horvat ansatz for BSk21 and SLy4, respectively, while the lower panels [(c), (d)] correspond to the Bowers-Liang ansatz for the same equations of state. Only those stellar configurations satisfying the stability and physical acceptability criteria discussed in Sec.~\ref{sec:equilib} are included.}
	\label{fig:freq-compactness_2x2}
\end{figure*}

Figure~\ref{fig:dt-mass_2x2} presents the damping time of the axial $w$-modes as a function of the gravitational mass for the same set of anisotropic stellar sequences shown in Fig.~\ref{fig:freq-mass_2x2}, namely for the BSk21 and SLy4 equations of state and for both anisotropy ansatzes. In all four panels, the damping time increases with mass along the stable branch. The curves are tightly clustered near $M \sim 1\,M_\odot$, where the damping time is nearly insensitive to the anisotropy strength at a fixed mass, whereas the separation among different anisotropy sequences becomes progressively more pronounced toward the high-mass end.

A clear dependence on the sign and magnitude of the anisotropy emerges at larger masses. For a fixed mass in the upper part of the sequence, configurations with positive anisotropy ($p_t>p_r$) typically exhibit shorter damping times than those with negative anisotropy ($p_r>p_t$). Consequently, the negative-anisotropy curves lie above the positive-anisotropy ones at high masses in each panel, and the disparity increases as one approaches the termination of the sequences. This behavior is particularly evident in the rapid upturn of the damping time near the endpoint of each curve, indicating that the damping becomes weaker close to the limiting configurations that satisfy the selection criteria adopted in Sec.~\ref{sec:equilib}.

Comparing the two anisotropy ansatzes, the Bowers--Liang results [panels (c) and (d)] span a broader mass range and reach substantially larger damping times than the corresponding Horvat results [panels (a) and (b)], consistent with the fact that the Bowers--Liang equilibrium sequences admit stable configurations up to higher masses for the parameter range considered. The EOS dependence is also evident: for a given ansatz and anisotropy strength, the BSk21 sequences extend to higher masses than those obtained with SLy4 and consequently attain larger damping times near their endpoints. As in the frequency--mass case, the termination of each curve reflects the subset of models that satisfy the stability and physical acceptability conditions adopted in Sec.~\ref{sec:equilib}, and therefore different anisotropy strengths terminate at different maximum masses.

Figure~\ref{fig:freq-compactness_2x2} shows the axial $w$-mode frequency as a function of the stellar compactness $M/R$ for anisotropic neutron-star sequences constructed using the BSk21 and SLy4 equations of state. In all four panels, the frequency exhibits an approximately linear dependence on compactness over the range of physically admissible configurations. This near-linearity persists across different anisotropy strengths and for both anisotropy ansatzes, indicating that compactness provides an efficient parametrization of the axial-mode spectrum.

\begin{figure*}[t]
	\centering
	
	\begin{subfigure}[t]{0.49\textwidth}
		\centering
		\includegraphics[width=\linewidth]{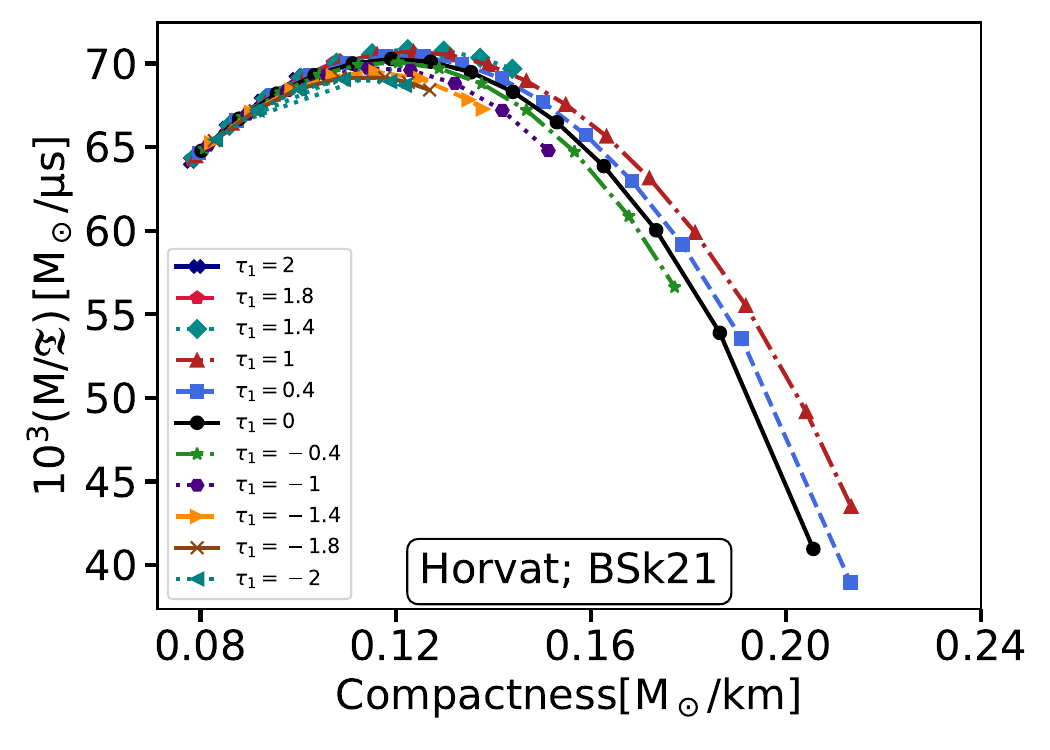}
		\caption{} \label{fig:dt_compactness_bsk21_hor}
	\end{subfigure}\hfill
	\begin{subfigure}[t]{0.49\textwidth}
		\centering
		\includegraphics[width=\linewidth]{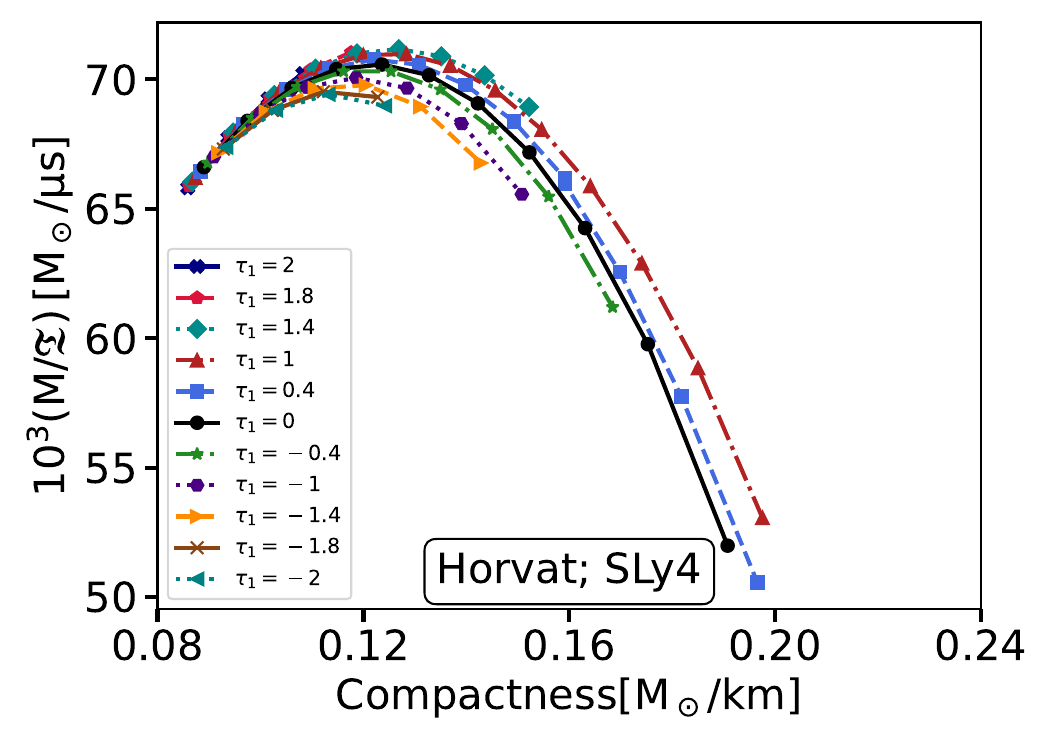}
		\caption{} \label{fig:dt_compactness_sly4_hor}
	\end{subfigure}
	
	\vspace{2mm}
	
	\begin{subfigure}[t]{0.49\textwidth}
		\centering
		\includegraphics[width=\linewidth]{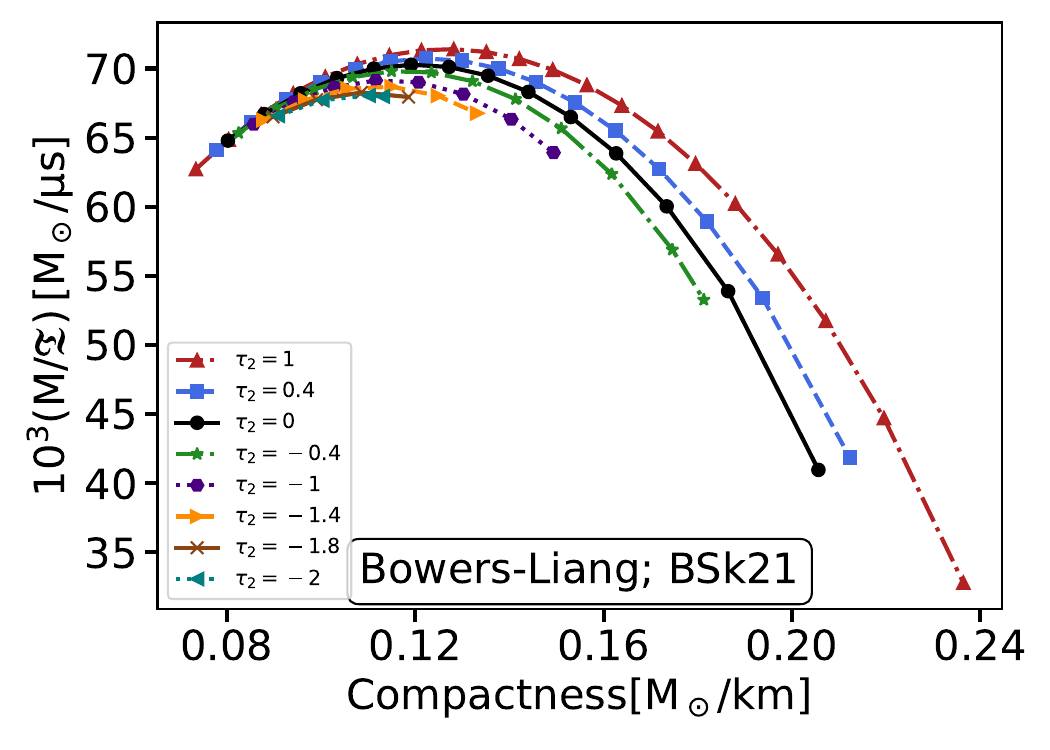}
		\caption{} \label{fig:dt_compactness_bsk21_bow}
	\end{subfigure}\hfill
	\begin{subfigure}[t]{0.49\textwidth}
		\centering
		\includegraphics[width=\linewidth]{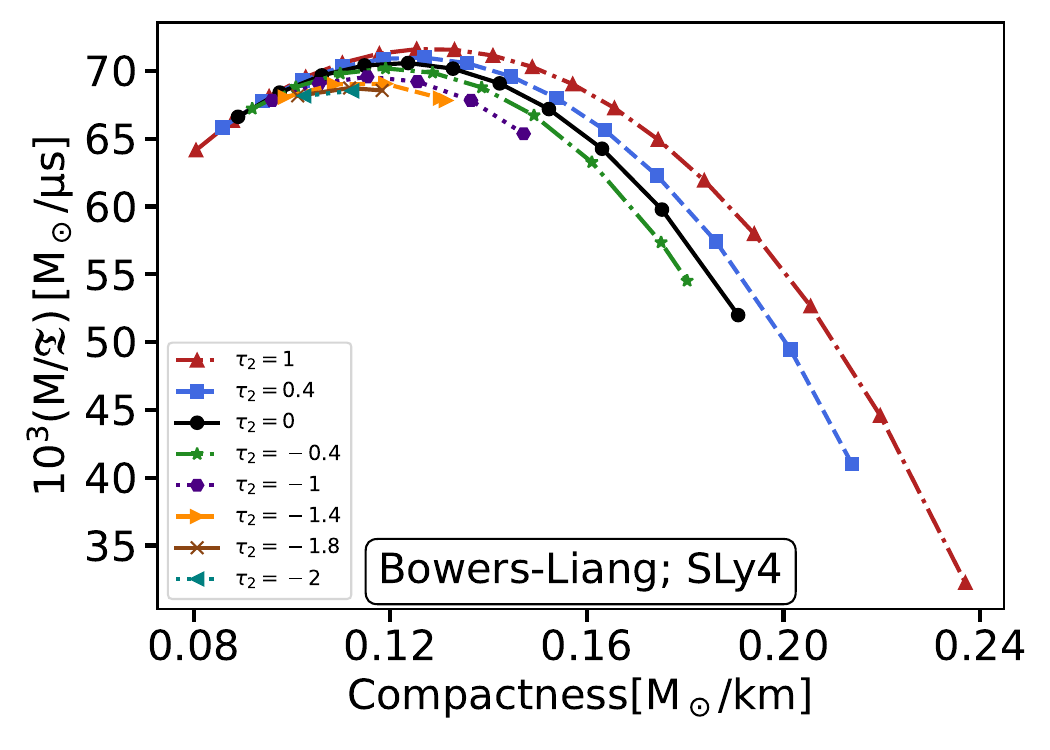}
		\caption{} \label{fig:dt_compactness_sly4_bow}
	\end{subfigure}
	
	\caption{Damping time-compactness relations of axial $w$-modes for anisotropic neutron-star models constructed with the BSk21 and SLy4 equations of state. The upper panels [(a), (b)] show results obtained using the Horvat ansatz for BSk21 and SLy4, respectively, while the lower panels [(c), (d)] correspond to the Bowers-Liang ansatz for the same equations of state. Only those stellar configurations satisfying the stability and physical acceptability criteria discussed in Sec.~\ref{sec:equilib} are included.}
	\label{fig:dt-compactness_2x2}
\end{figure*}

A closer inspection of the figure reveals that variations in the anisotropy parameter do not merely shift the $\mathcal{F}$--$M/R$ relation arbitrarily, but instead systematically modify both its intercept and its slope. For a given EOS and anisotropy ansatz, each value of the anisotropy parameter corresponds to a distinct, nearly straight line in the $\mathcal{F}$--$M/R$ plane. As the anisotropy is varied, these lines fan out, with changes in the slope becoming progressively more apparent toward larger compactness. This behavior motivates an empirical description of the form
\begin{equation}\label{emp_freq}
	\mathcal{F} \simeq A(\tau_i) + B(\tau_i)\,\frac{M}{R},
\end{equation}
where $\mathcal{F}$ is the axial $w$-mode frequency (in kHz), as defined above, and the coefficients $A$ and $B$ depend on the anisotropy parameter, with $\tau_i=\tau_1$ for the Horvat ansatz and $\tau_i=\tau_2$ for the Bowers--Liang ansatz.

For the Horvat ansatz [panels (a) and (b)], the dependence of both the slope and the intercept on $\tau_1$ is relatively weak at low and intermediate compactness, resulting in a tight clustering of the curves in this regime. The anisotropy-induced spread becomes noticeable only toward the upper end of the compactness range, where differences in the slopes lead to a clearer separation between sequences. In contrast, for the Bowers--Liang ansatz [panels (c) and (d)], the effect of anisotropy is more pronounced across the entire compactness interval considered. In this case, variations in $\tau_2$ lead to visibly different slopes as well as intercepts, producing a broader family of linear trends in the $\mathcal{F}$--$M/R$ plane.

The EOS dependence is comparatively mild when the frequency is expressed as a function of compactness: for a fixed ansatz and anisotropy strength, the BSk21 and SLy4 sequences follow similar linear trends, differing primarily in the maximum compactness reached before the termination of the stable branch. In the following, we quantify these trends by fitting the linear relation in Eq.~(\ref{emp_freq}) for each EOS and anisotropy strength, and we summarize the resulting fit coefficients.

To obtain explicit expressions for $A(\tau_i)$ and $B(\tau_i)$, we first compute the values of $\mathcal{F}$ for a range of stable neutron-star masses, from $1\,M_\odot$ up to the maximum stable mass, for several fixed values of $\tau_i$. For the Horvat ansatz, we consider $\tau_1$ in the range $-2 \leq \tau_1 \leq 2$ for both EOSs, while for the Bowers--Liang ansatz we take $\tau_2$ in the range $-2 \leq \tau_2 \leq 1$ for the same EOSs. For each value of $\tau_i$, we perform a linear fit between $\mathcal{F}$ and $M/R$, with the slope and intercept of each fit yielding the corresponding values of $B(\tau_i)$ and $A(\tau_i)$, respectively. In this procedure, we use $M$ in units of $M_\odot$, $R$ in kilometers, and $\mathcal{F}$ in kilohertz. We note that in geometrized units ($G = c = 1$), mass carries dimensions of length; accordingly, the solar mass serves as a unit of length, with $1\,M_\odot = 1.47664\,\mathrm{km}$, consistent with the adopted unit system.

Subsequently, for each of the four cases separately (two anisotropy ansatzes and, for each ansatz, two EOSs), we collect the fitted values of $A$, $B$, and the corresponding $\tau_i$, and perform polynomial fits of up to fourth order. The resulting empirical expression can be written as
\begin{equation}\label{emp:freq_expanded}
	\mathcal{F} \simeq \sum_{j = 0}^{4} \left(a_j \tau^j\right)\frac{M}{R}
	+ \sum_{j = 0}^{4} \left(b_j \tau^j\right) ,
\end{equation}
for all combinations of EOSs and anisotropy ansatzes considered above. In this expression, the coefficients $a_j$ and $b_j$ correspond to the polynomial representations of $B(\tau_i)$ and $A(\tau_i)$, respectively. The best-fit parameters for all cases are listed in Appendix~\ref{fit:param}.

To assess the quality of the fits, we compute the coefficient of determination, $\mathcal{R}^2$, defined as
\begin{equation}\label{cod}
	\mathcal{R}^2 = 1 - \frac{\sum_i (Y_{ni} - Y_{fi})^2}{\sum_i (Y_{ni} - \bar{Y})^2},
\end{equation}
where $Y_{ni}$ denotes the numerically computed value of the quantity $Y$ (here, the axial $w$-mode frequency), $Y_{fi}$ is the corresponding value obtained from the fit, and $\bar{Y}$ is the mean of the numerical data. Values of $\mathcal{R}^2$ exceeding $0.95$ are generally regarded as indicative of an excellent fit. For the Horvat ansatz, we obtain $\mathcal{R}^2 = 0.9944$ for BSk21 and $0.9977$ for SLy4, demonstrating excellent agreement in both cases. Similarly, for the Bowers--Liang ansatz, the corresponding values are $\mathcal{R}^2 = 0.9956$ for BSk21 and $0.9874$ for SLy4.

To examine the influence of anisotropy on the damping time, following Ref.~\cite{Blazquez-Salcedo:2018qyy} we plot the inverse of the damping time multiplied by the stellar mass, $10^3(M/\mathfrak{T})$, as a function of the compactness $M/R$ for various values of the anisotropy strength $\tau_i$. From the plots shown in Fig.~\ref{fig:dt-compactness_2x2}, it is evident that the relation follows an overall parabolic trend, with the location of the vertex and the curvature changing as the anisotropy strength is varied. This behavior motivates an empirical description of the form
\begin{equation}
	\frac{10^3 M}{\mathfrak{T}} = C(\tau_i)\left(\frac{M}{R}\right)^2
	+ D(\tau_i)\frac{M}{R}
	+ E(\tau_i) ,
\end{equation}
where $\mathfrak{T}$ is the axial $w$-mode damping time measured in $\mu$s, $M$ is expressed in units of $M_\odot$, and $R$ in kilometers. The coefficients $C$, $D$, and $E$ depend on the anisotropy parameter, with $\tau_i=\tau_1$ for the Horvat ansatz and $\tau_i=\tau_2$ for the Bowers--Liang ansatz.

We first compute the values of $10^3(M/\mathfrak{T})$ for a range of stable neutron-star masses, using the same mass range and anisotropy-parameter intervals adopted in the frequency analysis discussed above. For each fixed value of $\tau_i$, we then perform a parabolic fit between $10^3(M/\mathfrak{T})$ and the compactness $M/R$. The resulting fit coefficients $C$, $D$, and $E$ are obtained separately for each value of $\tau_i$. As in the previous subsection, we use $M$ in units of $M_\odot$, $R$ in kilometers, and $\mathfrak{T}$ in microseconds.

In the next step, for each of the four cases considered (two anisotropy ansatzes and, for each ansatz, two equations of state), we collect the fitted values of $C$, $D$, and $E$ together with the corresponding $\tau_i$ values and perform polynomial fits of up to sixth order. The resulting empirical expression can be written as
\begin{equation}\label{emp:freq_expanded}
	\frac{10^3 M}{\mathfrak{T}} \simeq
	\sum_{j = 0}^{6} \left(c_j \tau^j\right)\left(\frac{M}{R}\right)^2
	+ \sum_{j = 0}^{6} \left(d_j \tau^j\right)\frac{M}{R}
	+ \sum_{j = 0}^{6} \left(e_j \tau^j\right) ,
\end{equation}
for all combinations of equations of state and anisotropy ansatzes discussed above. In this expression, the coefficients $c_j$, $d_j$, and $e_j$ correspond to the polynomial representations of $C(\tau_i)$, $D(\tau_i)$, and $E(\tau_i)$, respectively. The best-fit parameters for all cases are listed in Appendix~\ref{fit:param}.

The quality of these fits is assessed using the coefficient of determination, $\mathcal{R}^2$, defined in Eq.~(\ref{cod}). For the Horvat ansatz with the BSk21 EOS, we obtain $\mathcal{R}^2 = 0.9679$, while for SLy4 the value is $0.9686$, indicating excellent agreement in both cases. Similarly, for the Bowers--Liang ansatz, the corresponding $\mathcal{R}^2$ values are $0.9994$ for BSk21 and $0.9842$ for SLy4.

\section{Summary and Conclusion}
\label{summ_con}

In this work, we have investigated the axial $w$-mode oscillations of anisotropic neutron stars. Equilibrium stellar configurations were constructed using two realistic equations of state, BSk21 and SLy4, together with two widely used prescriptions for modeling pressure anisotropy, namely the Horvat ansatz and the Bowers--Liang ansatz. For each case, we retained only those stellar models that satisfy the stability and physical acceptability criteria. On these backgrounds, the complex axial $w$-mode frequencies were computed by solving the perturbation equations using the continued-fraction method.

Our results show that the axial $w$-mode frequency decreases monotonically with increasing stellar mass for all anisotropic neutron-star sequences considered in this study. While this overall trend is common to both anisotropy prescriptions and both equations of state, the absolute value of the frequency and its rate of variation with mass depend sensitively on the nature and strength of the anisotropy. At relatively low stellar masses, configurations with dominant radial pressure ($p_r>p_t$) exhibit higher axial $w$-mode frequencies than those with dominant tangential pressure. Toward the upper end of the stable branch, however, this ordering is reversed, and configurations with $p_t>p_r$ attain higher frequencies at the same mass. These anisotropy-induced differences become increasingly pronounced near the high-mass termination of the stable sequences.

When expressed as a function of compactness, the axial $w$-mode frequency displays an approximately linear dependence on $M/R$, with anisotropy modifying both the slope and the intercept of the relation. For a given equation of state, variations in the anisotropy strength shift the frequency--compactness curves in a systematic manner, giving rise to a family of nearly linear relations. Compared to the Horvat prescription, the Bowers--Liang ansatz produces a broader spread in the frequency values over the allowed compactness range, reflecting the stronger influence of anisotropy in this model.

We have also examined the damping time associated with the axial $w$-modes. In all cases studied, the damping time increases with stellar mass, with a particularly rapid rise near the upper end of the stable branch. At a fixed mass, configurations with larger tangential pressure relative to radial pressure exhibit shorter damping times, while those with dominant radial pressure show significantly longer damping times. As in the case of the frequency, the sensitivity of the damping time to anisotropy is enhanced for more compact stars. In particular, neutron-star models constructed using the Bowers--Liang ansatz yield substantially larger damping times than those obtained with the Horvat prescription.

Motivated by the systematic trends observed in the numerical results, we have derived empirical expressions for both the axial $w$-mode frequency and the corresponding damping time as functions of stellar compactness and the anisotropy strength parameter. For each equation of state and anisotropy prescription, the frequency is well described by a linear function of compactness, with coefficients that depend polynomially on the anisotropy parameter. A similar empirical description is obtained for the damping time. The excellent quality of the fits across all cases considered indicates that these relations reliably capture the dependence of the axial $w$-mode spectrum on anisotropy within the physically admissible parameter space.

The results presented in this work demonstrate that pressure anisotropy leaves a clear and systematic imprint on the axial $w$-mode spectrum of neutron stars. Since axial modes are dominated by spacetime curvature and are efficient emitters of gravitational waves, these findings underscore the potential importance of anisotropy in gravitational-wave asteroseismology. If axial $w$-modes are detected in future observations, the empirical relations obtained here may provide a useful framework for constraining not only the neutron-star equation of state but also the nature and degree of pressure anisotropy in the stellar interior.

In future work, the present analysis can be extended by considering additional equations of state and exploring a broader range of anisotropy prescriptions. Such studies would allow the formulation of more general empirical relations for the frequency and damping time of axial $w$-mode oscillations in anisotropic neutron stars.

\section{ACKNOWLEDGMENTS}
The author thank the anonymous referee for constructive suggestions on the first version of the manuscript.

\section{DATA AVAILABILITY}
No observational data were created or analyzed in this study.

\appendix
\onecolumngrid

\section{Fitting coefficients for the axial $w$-mode frequency and damping time}\label{fit:param}

Tables~\ref{tab:fitparam1}--\ref{tab:fitparam4} list the fitting coefficients used to construct the empirical expressions for the axial $w$-mode frequency and damping time in Sec.~\ref{sec:numerical_results}.

\FloatBarrier

\begin{table}[H]
	\centering
	\caption{Fitting parameters for $A(\tau)$ for various equations of state and ansatzes.
		The coefficients $a_0$-$a_{4}$ are obtained from the fitting procedure described in Sec.~\ref{sec:numerical_results}.}
	\label{tab:fitparam1}
	\begin{tabular}{lccccc}
		\hline\hline
		EOS
		& $a_0$ & $a_1$ & $a_2$ & $a_3$ & $a_4$  \\
		\hline
		Horvat;BSk21        & -71.76 & 3.10 & -5.05 & -2.16 & 0.13\\
		Horvat;SLy4         & -74.40 & 1.26 & -1.74 & -1.41 & -0.23
		  \\
		Bowers--Liang;BSk21 & -72.69 & 6.40 & -1.53 & 1.22 & 0.70\\
		Bowers--Liang;SLy4  & -73.82 & 7.06 & -0.39 & 0.88 & 0.44\\
		\hline\hline
	\end{tabular}
\end{table}

\begin{table}[H]
	\centering
	
	\caption{Fitting parameters for $B(\tau)$ for various equations of state and ansatzes.
		The coefficients $b_0$-$b_{4}$ are obtained from the fitting procedure described in Sec.~\ref{sec:numerical_results}.}
	\label{tab:fitparam2}
	\begin{tabular}{lccccc}
		\hline\hline
		EOS
		& $b_0$ & $b_1$ & $b_2$ & $b_3$ & $b_4$  \\
		\hline
		Horvat;BSk21        & 23.97 & -0.65 & 0.58 & 0.22 & -0.03\\
		Horvat;SLy4         & 25.61 & -0.55 & 0.24 & 0.15 & 0.02\\
		Bowers--Liang;BSk21 & 24.08 & -1.94 & 0.02 & -0.16 & -0.08\\
		Bowers--Liang;SLy4  & 25.54 & -2.29 & -0.14 & -0.15 & -0.06\\
		\hline\hline
	\end{tabular}
\end{table}

\begin{table}[H]
	\centering
	\caption{Fitting parameters for $C(\tau)$ for various equations of state and ansatzes.
		The coefficients $c_0$-$c_{6}$ are obtained from the fitting procedure described in Sec.~\ref{sec:numerical_results}.}
	\label{tab:fitparam3}
	\begin{tabular}{lccccccc}
		\hline\hline
		EOS
		& $c_0$ & $c_1$ & $c_2$ & $c_3$ & $c_4$ & $c_5$
		& $c_6$  \\
		\hline
		Horvat;BSk21        & -3835.37 & 251.39 & 328.62 & 62.42 & -134.92 & -16.64 & 15.09\\
		Horvat;SLy4         & -3886.99 & 409.76 & 102.33 & -20.63 & -41.88 & -2.42 & 2.46\\
		Bowers--Liang;BSk21 & -3871.76 & 575.89 & 427.63 & -51.13 & -226.24 & -53.41 & 6.27\\
		Bowers--Liang;SLy4  & -3974.65 & 694.86 & 131.44 & -374.77 & -86.47 & 203.80 & 77.64\\
		\hline\hline
	\end{tabular}
\end{table}

\begin{table}[H]
	\centering
	\caption{Fitting parameters for $D(\tau)$ for various equations of state and ansatzes.
		The coefficients $d_0$-$d_{6}$ are obtained from the fitting procedure described in Sec.~\ref{sec:numerical_results}.}
	\label{tab:fitparam4}
	\begin{tabular}{lccccccc}
		\hline\hline
		EOS
		& $d_0$ & $d_1$ & $d_2$ & $d_3$ & $d_4$ & $d_5$
		& $d_6$  \\
		\hline
		Horvat;BSk21        & 920.24 & -29.69 & -81.79 & -19.01 & 31.40 & 4.46 & -3.48\\
		Horvat;SLy4         & 946.96 & -72.88 & -30.86 & 4.38 & 11.52 & 0.57 & -0.86\\
		Bowers--Liang;BSk21 & 929.38 & -86.25 & -108.59 & 4.38 & 56.32 & 17.07 & -0.39\\
		Bowers--Liang;SLy4  & 970.13 & -120.11 & -37.06 & 91.32 & 23.75 & -50.33 & -19.77\\
		\hline\hline
	\end{tabular}
\end{table}

\begin{table}[H]
	\centering
	
	\caption{Fitting parameters for $E(\tau)$ for various equations of state and ansatzes.
		The coefficients $e_0$-$e_{6}$ are obtained from the fitting procedure described in Sec.~\ref{sec:numerical_results}.}
	\label{tab:fitparam4}
	\begin{tabular}{lccccccc}
		\hline\hline
		EOS
		& $e_0$ & $e_1$ & $e_2$ & $e_3$ & $e_4$ & $e_5$
		& $e_6$  \\
		\hline
		Horvat;BSk21        & 15.32 & 0.61 & 4.75 & 1.27 & -1.75 & -0.28 & 0.19\\
		Horvat;SLy4         & 13.01 & 3.29 & 2.03 & -0.22 & -0.73 & -0.03 & 0.06\\
		Bowers--Liang;BSk21 & 14.78 & 3.31 & 6.37 & 0.06 & -3.31 & -1.16 & -0.02\\
		Bowers--Liang;SLy4  & 11.55 & 5.38 & 2.32 & -5.43 & -1.54 & 3.01 & 1.21\\
		\hline\hline
	\end{tabular}
\end{table}

\FloatBarrier

\twocolumngrid

\bibliography{main}

\end{document}